\documentclass[10pt,letterpaper]{article}

\usepackage{setspace}

\usepackage{ccn}
\usepackage{pslatex}
\usepackage{apacite}
\usepackage{natbib}
\usepackage{lineno}

\usepackage{amsmath}
\usepackage{graphicx}

\usepackage[utf8]{inputenc} 
\usepackage[T1]{fontenc}    
\usepackage{booktabs}       
\usepackage{amsfonts, amsmath}       
\usepackage{nicefrac}       
\usepackage{microtype}      
\usepackage{xcolor}         
\usepackage{enumitem}

\usepackage{bbm}

\usepackage{caption}
\usepackage{subcaption}
\usepackage{adjustbox}
\usepackage{wrapfig}
\usepackage{float}

\usepackage{multirow}

\usepackage[hidelinks]{hyperref}       

\makeatletter
\def\subsubsection{\@startsection{subsubsection}{3}{\z@}%
  {-1.2ex plus -0.2ex minus -0.2ex}%
  {0.6ex plus 0.2ex}%
  {\normalsize\bfseries\raggedright}}
\makeatother

\title{Geometric organization of olfactory descriptor data in the Poincaré disk}
 
\author{{\large \bf Aniss Aiman Medbouhi} \\
  Department of Robotics, Perception and Learning\\
  School of Electrical Engineering and Computer Science\\
KTH Royal Institute of Technology, Sweden
  \AND {\large \bf Farzaneh Taleb} \\
Department of Robotics, Perception and Learning\\
School of Electrical Engineering and Computer Science\\
KTH Royal Institute of Technology, Sweden
  \AND {\large \bf Giovanni Luca Marchetti} \\
Department of Mathematics\\
School of Engineering Sciences\\
KTH Royal Institute of Technology, Sweden
  \AND {\large \bf Danica Kragic} \\
Department of Robotics, Perception and Learning\\
School of Electrical Engineering and Computer Science\\
KTH Royal Institute of Technology, Sweden
}

\begin{document}


\maketitle

\section*{ORCID iDs}

Aniss Aiman Medbouhi: 000-0002-6649-3325\\
Farzaneh Taleb: 0000-0003-4482-1460\\
Giovanni Luca Marchetti 0009-0004-8248-229X\\
Danica Kragic: 0000-0003-2965-2953

\section*{Correspondence to be sent to}

Address: Lindstedtsvägen 24, 114 28 Stockholm, Sweden.\\
Email: medbouhi@kth.se

\clearpage


\section*{Abstract}
{
\bf
Odor quality is commonly represented using high dimensional descriptor profiles, yet their low dimensional organization remains unclear. We investigated whether a two-dimensional hyperbolic embedding can provide an interpretable representation of this structure. We applied hyperbolic metric multidimensional scaling to two complementary datasets: 480 Sagar rating profiles from three participants rating 160 odorants on 15 continuous descriptors, and 4983 GoodScents--Leffingwell molecules annotated with 138 binary descriptors. The embeddings substantially preserved pairwise descriptor distances, supporting subsequent analyses of radial and angular organization. In Sagar, \emph{rating profile entropy} was strongly and negatively associated with hyperbolic radius (Pearson \(r=-0.77\pm0.05\)), with diffuse profiles closer to the center and concentrated profiles closer to the boundary. This radial organization emerged primarily at the level of the full descriptor profile, rather than any individual descriptor, and remained robust across alternative descriptor representations, participant specific analyses, and averaged ratings. \emph{Sweet}, \emph{musky}, \emph{fruity}, and \emph{pleasantness} showed the strongest directional trends (mean \(R^2=0.40\) to \(0.51\)). In GoodScents--Leffingwell, \emph{active label entropy}, reflecting descriptor multiplicity, increased with radius (\(r=0.87\pm0.03\)), whereas \emph{orthogonalized descriptor entropy}, reflecting spread across orthogonal modes, decreased with radius (\(r=-0.88\pm0.01\)). Related binary descriptors occupied coherent localized high-density regions. These findings reveal complementary radial and angular organization in the hyperbolic representation of olfactory descriptor data. They support hyperbolic mapping as an interpretable descriptive framework in which radius summarizes global profile properties, while the angular component captures continuous descriptor gradients and categorical organization.
}

\begin{quote}
\small
\textbf{Keywords:} 
olfactory perception, hyperbolic geometry, dimensionality reduction. 
\end{quote}

\section{Introduction}

Human odor perception is challenging to organize within a simple coordinate system. While substantial progress has been made in characterizing visual and auditory perception through mathematical models and structured representations~\citep{sucholutsky2023getting, brohan2023rt, du2022survey, ganis2004brain, friederici2012cortical}, olfaction lacks a comparable theoretical framework. Unlike vision, where color perception has been systematically mapped through the Commission Internationale \cite{commission1931commission} color spaces, or audition, where auditory signals can be systematically represented in the frequency domain~\citep{evans1977frequency}, no comparably established and widely accepted mapping exists for the olfactory perceptual space. Odor quality does not vary along a single dominant physical continuum, and similar odor percepts can arise from chemically diverse molecules. Perceptual descriptions also depend on the odorants presented and their concentration, the response task, the available vocabulary, and individual and cultural differences in perceptual and verbal strategies \citep{kaeppler2013odor, Doty2025Odors}. Consequently, in the present study, we treat an olfactory perceptual space as a representation of similarities and differences among odor percepts or descriptor profiles, rather than as a direct and universal mapping from molecular structure to subjective experience.

A long tradition of olfactory research has used similarity judgments, sorting tasks, descriptor ratings, factor analysis, and multidimensional scaling to characterize these relationships. Such studies have identified broad hedonic and semantic trends, but they have not produced a single agreed dimensional organization of odor quality \citep{schiffman1974PhysicochemicalDimensionsOdor,MADANYMAMLOUK2004DimensionOlfactoryPerceptionSpace,koulakov2011search,Magnasco2015DimensionOlfactorySpace,Meister2015DimensionalityOdorSpace}. The resulting maps depend on the experimental setup, stimulus set, the descriptors, the participants, and the analysis method \citep{kaeppler2013odor}. A recent computational work has constructed learned odor representations from molecular and perceptual data, namely Principal Odor Map \citep{POM_PrincipalOdorMap_Lee2023}, while recent taxonomy based approaches have explicitly examined hierarchical relations among odor descriptors \citep{Sajan2026HierarchiesSmell}. These developments provide powerful representations for prediction, but the geometric structure of descriptor based odor spaces remains poorly understood.

Most low dimensional representations of odor perception have been formulated in Euclidean space. Hyperbolic geometry offers an alternative representation in which the amount of available space increases exponentially with distance from the origin. This property makes hyperbolic spaces effective for representing data with branching or hierarchical organization \citep{Sarkar2012_LowDistortionEmbeddingTreesHyperbolicPlane, nickel2017poincare, Klimovskaia2020SingleCellPoincareMap, Zhou2021HyperbolicGeometryGeneExpression, Zhang_2022_BrainSpatialRepresentationHyperbolic}. In olfaction, \citet{Zhou2018HyperbolicGeometryOlfactorySpace} reported that the statistics of natural odor mixtures and human perceptual descriptions were compatible with hyperbolic geometry. However, it remains unclear whether an interpretable two-dimensional hyperbolic representation can preserve relationships among odor descriptor profiles, retain established perceptual dimensions such as pleasantness \citep{crocker1927analysis, Khan2007PredictingOdorPleasantness, koulakov2011search, snitz2013predicting, licon2018pleasantness} as interpretable directions, and reveal how global profile properties are organized along its radial coordinate. It is also unknown whether comparable geometric organization appears across continuous participant ratings and large binary descriptor databases.
Here, we investigate these questions by employing a hyperbolic version of metric multidimensional scaling (MDS) \citep{torgerson1952MDS, kruskal1964multidimensional, Walter2004HyperbolicMDS, Sala2018_RepresentationTradeoffsHyperbolicEmbeddings, keller2020hydra}. This method represents each observation as a point in the two-dimensional Poincaré disk by minimizing differences between pairwise distances in the original descriptor space and the corresponding pairwise hyperbolic distances in the embedding. This choice is not intended as an estimate of the intrinsic dimensionality of olfactory perception, nor as evidence that neural olfactory representations are themselves two-dimensional or hyperbolic. Rather, the two-dimensional representation facilitates both visualization and interpretation by allowing center to boundary variation to be distinguished from angular organization.

We analyze two complementary datasets. The Sagar dataset~\citep{sagar2023high} contains continuous perceptual ratings for monomolecular odorants from three participants, allowing us to examine graded descriptor profiles, organization within individual participants, and participant averaged ratings. The GoodScents--Leffingwell dataset \citep{OpenPOM} contains binary expert annotations for several thousand molecules and provides a larger scale representation of how odor descriptors are jointly assigned across molecules. Using these datasets, we first verify pairwise distance preservation and then address three main questions: (i) Are global properties of descriptor profiles associated with hyperbolic radius?; (ii) Do continuous descriptors show directional organization and binary descriptors occupy localized coherent regions?; (iii) Are the observed relationships robust across random initializations, alternative descriptor representations, and where available, individual and averaged ratings? Statistical significance is assessed using restricted permutation tests adapted to the structure of each dataset.

We previously presented preliminary analyses of the Sagar
dataset in conference abstracts \citep{Taleb2025TowardsDiscoveringHierarchyOlfactoryHyperbolic,
Medbouhi2025CCN}. These preliminary contributions used a hyperbolic contrastive learning objective optimized with Riemannian stochastic gradient descent and identified an initial association between descriptor profile entropy and hyperbolic radius. The present study substantially extends this dataset-specific observation into a broader and statistically validated framework for interpreting hyperbolic olfactory representations. We replace the contrastive objective with
hyperbolic metric multidimensional scaling and Riemannian Adam optimization, which enables application to the substantially larger and binary GoodScents--Leffingwell dataset. We further introduce an
explicit radial--angular decomposition, evaluate pairwise distance preservation, develop complementary hyperbolic visualizations of continuous descriptor directions and binary descriptor regions, and assess the resulting organization through restricted permutation tests and extensive robustness analyses.

The results reveal complementary radial and angular organization. In the Sagar data, \emph{rating profile entropy} is strongly and negatively associated with radius, such that diffuse descriptor profiles are located closer to the center and more concentrated profiles closer to the boundary. This relationship is stronger than the radial association of any individual descriptor and persists across alternative descriptor representations, participant specific analyses, and averaged ratings. Several continuous descriptors, including sweet, musky, fruity, and pleasantness, show consistent \emph{directional trends}. In the GoodScents--Leffingwell data, radius is also associated with entropy related properties of the descriptor profiles, although the direction and interpretation of these relationships depend on how entropy is defined. Related binary odor descriptors additionally occupy coherent regions of the disk. Together, these findings support two-dimensional hyperbolic mapping as a descriptive framework for separating global descriptor profile organization from descriptor specific gradients and categorical odor quality structure.

\section{Materials and methods}

\subsection{Study overview}

We first construct descriptor vectors for each observation in the two datasets. We then learn a two-dimensional hyperbolic embedding that preserves pairwise distances between these vectors. The learned representation is evaluated in two stages. First, we quantify how well hyperbolic distances preserve the input descriptor geometry. Second, we test whether interpretable variables are organized along the radial and angular components of the Poincaré disk. The radial analyses examine entropy and continuous descriptor ratings. The angular analyses use tangent space directional trends for continuous ratings and hyperbolic density regions for binary annotations.

\subsection{Data sources} \label{sec: datasets}

\paragraph{Sagar.}

We used the publicly available \emph{Sagar} dataset~\citep{sagar2023high}, obtained from the Pyrfume repository~\citep{hamel2024pyrfume}. The original study collected perceptual ratings for $160$ unique monomolecular odor stimuli per subject, across $S=3$ subjects. In our union dataset, each subject contributes $160$ subject--odorant observations, giving $N=480$ observations in total. Because the odor sets are not identical across subjects, these $480$ observations correspond to $195$ unique Compound Identifiers (CIDs): $125$ CIDs are observed for all three subjects, $35$ CIDs are observed only for subject 1, and $35$ CIDs are observed for subjects 2 and 3.

Each subject rated odors using $18$ perceptual descriptors. $15$ descriptors are common to all subjects: intensity, pleasantness, fishy, burnt, sour, decayed, musky, fruity, sweaty, cool, floral, sweet, warm, bakery, and spicy. The remaining descriptors are subject-specific and are therefore excluded from our analysis to ensure a common descriptor space across subjects. All ratings were normalized to the range $[-1,1]$, yielding descriptor vectors in $[-1,1]^{15}$. The diversity of odorants and perceptual attributes makes this dataset well suited for investigating the human odor perception.

\paragraph{GoodScents--Leffingwell.}

We also used the GoodScents--Leffingwell (GSLF) dataset distributed with OpenPOM~\citep{OpenPOM}, which combines odor annotations from the \cite{gs} company and \cite{lf} databases following the curation procedure of~\citet{lee2022principal}. The resulting dataset contains $4983$ molecules annotated with $138$ expert-defined odor descriptors, such as \emph{fruity}, \emph{jasmin}, and \emph{leathery}. Unlike the Sagar dataset, these annotations are binary labels rather than continuous ratings: each descriptor is either present or absent for a given molecule. Thus, each molecule is represented by a multi-label binary vector $b_i \in \{0,1\}^{138}$.

Each molecule appears once in the dataset, so there is no subject-level or repeated-measures structure. This makes GSLF complementary to Sagar: Sagar provides continuous subject-specific perceptual ratings for a smaller set of odorants, whereas GSLF provides a larger-scale binary descriptor representation over thousands of molecules. We use GSLF to test whether radial entropy organization and descriptor-region structure also appear in a large expert annotated odor dataset.

\subsection{Hyperbolic metric multidimensional scaling} \label{sec: hMDS}

\paragraph{Poincaré model.} Hyperbolic geometry is a non Euclidean geometry with constant negative curvature. We employ the Poincaré disk model to embed perceptual descriptor data: the goal is to learn a two-dimensional hyperbolic representation that preserves pairwise input distances. The Poincaré disk $\mathbb{P}^{2}$ represents the two-dimensional hyperbolic space inside the open Euclidean unit disk,
\[
\mathbb{P}^{2}=\{z\in\mathbb{R}^{2}:\lVert z\rVert<1\},
\]
where $\|\cdot\|$ denotes the Euclidean norm. Although the disk is drawn in Euclidean coordinates, distances are measured using the hyperbolic metric. For points $z_1,z_2\in\mathbb{P}^{2}$, the hyperbolic distance is
\[
d_{\mathbb{P}}(z_1,z_2)=\operatorname{arcosh}\!\left(1+\frac{2\lVert z_1-z_2\rVert^{2}}{(1-\lVert z_1\rVert^{2})(1-\lVert z_2\rVert^{2})}\right).
\]
Distances therefore expand near the boundary of the disk. This allows many mutually separated observations to be represented at large radii, which is useful when the data contain branching or hierarchical structure.

\paragraph{Radial and angular organization.} The two-dimensional representation also provides a useful descriptive decomposition. The hyperbolic radius $d_{\mathbb{P}}(z,0)$ measures center to boundary position, whereas angular position describes direction around the disk. Neither coordinate has an intrinsic perceptual meaning. We establish empirically their interpretation by testing associations with entropy and descriptor values. In addition, the absolute angular orientation can rotate or reflect across model initializations, so only relative directional and regional organization is interpreted.

\paragraph{Objective function.} Given perceptual descriptor vectors $\{x_i\}_{i=1}^{N}$, we compute input distances $D_{ij}^{\mathrm{in}}=\lVert x_i-x_j\rVert$. We learn points $\{z_i\}_{i=1}^{N}\subset\mathbb{P}^{2}$, called the \emph{embeddings}, by minimizing the following loss:
\[
\mathcal{L}=\frac{1}{N^{2}}\sum_{i,j=1}^{N}\left(D_{ij}^{\mathrm{in}}-D_{ij}^{\mathrm{emb}}\right)^{2},
\]
where $D_{ij}^{\mathrm{emb}} = d_{\mathbb{P}}(z_i,z_j)$ are the hyperbolic embedding distances. This is the metric multidimensional scaling principle expressed with hyperbolic distances \citep{torgerson1952MDS,kruskal1964multidimensional,Sala2018_RepresentationTradeoffsHyperbolicEmbeddings,keller2020hydra}.

\paragraph{Optimization.}
The embedding coordinates are initialized in the Poincaré disk via a hyperbolic analog of Gaussian distribution \citep{Nagano2019wrapped}, and optimized by minimizing the loss ${L}$ with Riemannian Adam \citep{becigneul2019riemannian}. Full geometric expressions, optimization updates, numerical stability procedures and implementation details are provided in the Supplementary material~\ref{app:hyperbolic_details}.

\subsection{Evaluation of embedding quality and geometric organization}

We evaluate the learned embeddings in two steps. First, we use distance preservation as a quality control measure to verify that the hyperbolic metric MDS optimization has produced embeddings that preserve the input descriptor geometry. Second, we analyze the structure of the learned embeddings by asking how entropy, continuous descriptor ratings, and binary descriptor labels are organized in the Poincaré disk. These analyses are divided into radial organization, and angular organization comprising directional trends for continuous descriptors and high density regions for binary descriptors.

\paragraph{Embedding quality.}

As a quality control step, we evaluate whether the learned hyperbolic embeddings preserve the pairwise geometry of the input descriptor space. This check is important because the subsequent radial and directional analyses are meaningful only if the embedding retains the main distance structure of the original data. For each training configuration, we compute the fixed input distance matrix $D^{in}$. For each random seed $m$, we compute the corresponding hyperbolic embedding distance matrix $D^{emb,m}$. We then vectorize the upper triangular entries of both matrices, excluding the diagonal, and compute Pearson and Spearman correlations between the two distance vectors. Pearson correlation measures linear agreement between input and embedding distances, whereas Spearman correlation measures preservation of the distance ranking. We report mean and standard deviation over random seeds. We use these correlations as embedding quality metrics, not as hypothesis tests, and therefore do not report analytic p-values for them.

\subsubsection{Entropy}

Entropy is a classical quantity in thermodynamics, statistical physics, and information theory, where it is used to quantify disorder, uncertainty, or the spread of a probability distribution. In order to evaluate and analyze the structure of the inferred hyperbolic embedding, we propose to employ entropy as a scalar summary of descriptor organization for each observation. The two datasets contain different types of descriptor values, so entropy has to be defined through a dataset specific probability vector. In the Sagar dataset, descriptors are continuous ratings, and entropy summarizes the spread of graded descriptor strengths. In the GSLF dataset, descriptors are binary annotations, and entropy summarizes the multiplicity of active odor labels. In robustness analyses, we also compute entropy after transforming descriptor vectors into an orthogonalized representation, which tests whether radial entropy organization depends on the orthogonal modes of variation.

For each observation $i$, we construct a probability vector
\[
p_i = (p_{i,1},\ldots,p_{i,n}),
\]
defined over the $n$ descriptors, and compute
\[
H_i
=
-\sum_{k=1}^{n} p_{i,k}\log p_{i,k}.
\]
This common entropy formula is used for both datasets, but the construction of $p_i$ differs according to the data type. This distinction is important because the resulting entropy measures answer related but not identical questions.

\paragraph{Rating profile entropy.}
For continuous descriptor ratings, as in the Sagar dataset, each observation is represented by a vector $x_i \in \mathbb{R}^n$, where $x_{i,k}$ is the rating of descriptor $k$ for observation $i$. Since these ratings can be positive or negative after normalization, we convert them into a probability distribution using a softmax:
\[
p_{i,k}
=
\frac{\exp(x_{i,k})}
{\sum_{\ell=1}^{n}\exp(x_{i,\ell})}.
\]
We refer to the resulting $H_i$ quantity as \textbf{rating profile entropy}. It is high when the continuous descriptor profile is diffuse across many descriptors, and low when the profile is concentrated on one or a few descriptors. Therefore, in the Sagar dataset, the radial analysis asks whether odors with diffuse or ambiguous continuous rating profiles are placed differently in the hyperbolic embedding from odors with more concentrated descriptor profiles.

\paragraph{Active label entropy.}
For binary descriptor annotations, as in the GSLF dataset, each molecule is represented by a vector $b_i \in \{0,1\}^n$, where $b_{i,k}=1$ if descriptor $k$ is assigned to molecule $i$, and $b_{i,k}=0$ otherwise. Applying the same softmax construction directly to the binary vector would give positive probability to inactive labels, because $\exp(0)>0$. This would make absent descriptors contribute to the entropy, which is not the intended interpretation of a binary descriptor annotation. We therefore define the probability distribution only over the active descriptor set. Let
\[
A_i = \{k : b_{i,k}=1\}
\]
be the set of active descriptors for molecule $i$, and let
\[
K_i = |A_i|
\]
be the number of active descriptors. For molecules with at least one active descriptor, we define
\[
p_{i,k}
=
\begin{cases}
1/K_i, & \text{if } k \in A_i,\\
0, & \text{otherwise}.
\end{cases}
\]
With this definition, the entropy reduces to
\[
H_i
=
-\sum_{k=1}^{n} p_{i,k}\log p_{i,k}
=
\log K_i.
\]
Molecules with no active descriptor have undefined entropy and are excluded from entropy based analyses. Thus, in the GSLF dataset, entropy measures the breadth or multiplicity of the expert descriptor profile. We refer to this quantity as \textbf{active label entropy}. It is high when many descriptors are assigned to a molecule, and low when only one or a few descriptors are assigned.

Importantly, \emph{active label entropy} is not equivalent to the \emph{rating profile entropy} used for continuous descriptor ratings. A molecule can have many active binary labels and therefore high \emph{active label entropy}, while a hypothetical continuous rating profile over the same descriptors could still be concentrated on one or a few dominant qualities and therefore have low \emph{rating profile entropy}. Thus, \emph{active label entropy} measures descriptor multiplicity, whereas \emph{rating profile entropy} measures the spread of graded descriptor strengths. The sign of the radius entropy relationship should therefore be interpreted relative to the entropy definition used in each dataset.

\paragraph{Orthogonalized descriptor entropy.}
For robustness analysis, we express the descriptor matrix in an orthogonal coordinate system, so that the resulting dimensions no longer correspond to individual descriptor ratings or labels but to orthogonal modes of variation in the descriptor data. We then construct $p_i$ by applying a softmax, and refer to the resulting quantity as \textbf{orthogonalized descriptor entropy}. This entropy asks whether the transformed descriptor profile of an observation is balanced across several orthogonal modes of variation, or whether it is dominated by one or a few modes. A high \emph{orthogonalized descriptor entropy} indicates a more diffuse profile across modes, whereas a low value indicates a more concentrated one. This entropy is interpreted as a robustness measure rather than as a direct measure of descriptors profile or multiplicity.

Overall, these entropy measures ask related but distinct questions. \emph{Rating profile entropy} tests whether diffuse graded descriptor profiles are organized along the hyperbolic radius. \emph{Active label entropy} tests whether molecules associated with many odor qualities are organized radially. \emph{Orthogonalized descriptor entropy} tests whether radial entropy organization persists after replacing the original descriptor basis by orthogonal modes of variation.

\subsubsection{Geometric organization.}

\paragraph{Radial organization.}

We evaluate whether scalar quantities associated with observations are organized along the radial coordinate of the learned hyperbolic embedding. For each random seed $m=1,\ldots,M$, let $z_i^{(m)} \in \mathbb{P}^2$ denote the learned embedding of observation $i$. Its hyperbolic radius is defined as
\[
r_i^{(m)} = d_{\mathbb{P}}(z_i^{(m)},0).
\]
Let $y_i$ be a scalar variable associated with observation $i$. Depending on the analysis, $y_i$ can be an entropy value, or a continuous descriptor rating. For each seed $m$, we compute the signed Pearson correlation
\[
c_m(y) = \mathrm{corr}(r^{(m)},y),
\]
where $r^{(m)}$ is the vector of radii and $y$ is the vector of scalar values for the corresponding observations. We also compute the Spearman rank correlation as a non-parametric robustness measure. We report mean and standard deviation over random seeds.

In the Sagar union dataset, each observation $i$ corresponds to a subject--odorant pair $(o,s)$, with subject $s\in\{1,2,3\}$. In this case, the generic index $i$ can be identified with $(o,s)$. For entropy, the scalar value is $y_{(o,s)}=H_{(o,s)}$. For descriptor $k$, the scalar value is $y_{(o,s)}=x_{(o,s),k}$.

In the GSLF dataset, each observation corresponds to one annotated molecule. For radial analysis, the scalar variable is an entropy value $y_{i}=H_{i}$.

For the radial permutation test, the observed statistic associated with $y$ is the mean Pearson correlation over seeds:
\[
T_{\mathrm{rad},\mathrm{obs}}(y) =
\frac{1}{M}
\sum_{m=1}^{M}
c_m(y).
\]
The sign of the radial correlation indicates whether the scalar quantity tends to increase or decrease toward the boundary.

\paragraph{Tangent space representation.}

To characterize directional variation in continuous descriptor ratings, we use the tangent space of the Poincaré disk at the origin. The tangent space $T_0\mathbb{P}^2$ is a two-dimensional Euclidean vector space that provides a linear coordinate system centered at the origin of the disk. This allows standard linear regression to be applied to the embedded observations while retaining their radial and directional organization.

For an embedded point $z \in \mathbb{P}^2$, its tangent space coordinate is obtained using the logarithmic map at the origin:
\[
\log_0(z)
=
\begin{cases}
\operatorname{artanh}\left(\|z\|\right)\dfrac{z}{\|z\|},
& z \neq 0,\\[0.6em]
0,
& z=0.
\end{cases}
\]
This transformation preserves the direction of the point from the origin while expressing its position in a linear coordinate system.

For visualization, a tangent vector $u \in T_0\mathbb{P}^2$ can be mapped back to the Poincaré disk using the exponential map at the origin:
\[
\exp_0(u)
=
\begin{cases}
\tanh(\|u\|)\dfrac{u}{\|u\|},
& u \neq 0,\\[0.6em]
0,
& u=0.
\end{cases}
\]
The exponential and logarithmic maps are inverses of one another at the origin. General expressions and additional geometric operations used during optimization are provided in the Supplementary material \ref{app:hyperbolic_details}.

\paragraph{Angular organization of continuous descriptors: directional trends.}

For continuous descriptor ratings, we analyze directional trends in the learned Poincaré disk. This analysis is applied to the Sagar dataset, where descriptor values are graded ratings. While the radial analysis captures center to boundary variation, the directional analysis captures whether a descriptor changes primarily along a dominant angular direction in the embedding. We therefore fit, for each descriptor, a best-fitting plane over the tangent-space coordinates of the embedded points, and consider the direction in which this plane increases most steeply as a global summary of the descriptor's directional trend.

More formally, for each seed $m$, we first map the embedded points to the tangent space at the origin:
\[
u_{\left(o,s\right)}^{\left(m\right)} = \log_0\left(z_{\left(o,s\right)}^{\left(m\right)}\right).
\]
The tangent space $T_0\mathbb{P}^2$ is identified with $\mathbb{R}^2$. For each descriptor $k$, we fit a linear regression in this tangent space:
\[
x_{\left(o,s\right),k} \approx \beta_{k,m}^T u_{\left(o,s\right)}^{\left(m\right)} + b_{k,m}.
\]
Here, $x_{o,s,k}$ is the value of descriptor $k$ for odorant $o$ and subject $s$. The parameters $\beta_{k,m} \in \mathbb{R}^2$ and $b_{k,m} \in \mathbb{R}$ are obtained by minimizing the squared error:
\[
\sum_{\left(o,s\right)\in \mathcal{I}}
\left(
x_{\left(o,s\right),k} - \beta_{k,m}^T u_{\left(o,s\right)}^{(m)} - b_{k,m}
\right)^2.
\]
The vector $\beta_{k,m}$ gives the direction in the tangent space along which the fitted descriptor $k$ increases most strongly under this global linear approximation. For visualization, we normalize this direction as
\[
v_{k,m} =
\frac{\beta_{k,m}}{\|\beta_{k,m}\|},
\]
and map the corresponding vector back to the Poincaré disk using the exponential map:
\[
\gamma_{k,m}(t) = \exp_0\left(t v_{k,m}\right).
\]
$\gamma_{k,m}(t)$ is represented as an arrow indicating the direction of steepest increase in the rating of descriptor $k$, and $t$ only controls the displayed arrow length. Since the absolute orientation of the embedding can rotate or reflect across random seeds, these arrows are used only for visualization of representative embeddings. Their displayed lengths are arbitrary and do not represent the strength of the directional trend, which is quantified separately using the coefficient of determination $R^2$.

For descriptor $k$ and seed $m$, let $\hat{x}_{o,s,k}^{(m)}$ be the value predicted by the fitted linear model. We define
\[
R_{k,m}^2 =
1 -
\frac{
\sum_{\left(o,s\right)\in\mathcal{I}}
\left(
x_{\left(o,s\right),k} - \hat{x}_{\left(o,s\right),k}^{(m)}
\right)^2
}{
\sum_{\left(o,s\right)\in\mathcal{I}}
\left(
x_{\left(o,s\right),k} - \bar{x}_k
\right)^2
}.
\]
Here, $\bar{x}_k$ is the mean value of descriptor $k$ over all subject--odorant observations. A high value of the coefficient of determination $R_{k,m}^2$ indicates that descriptor $k$ is well summarized by a global directional trend in the tangent-space representation. We report the mean and standard deviation of $R_{k,m}^2$ over random seeds.

For the angular permutation test, the observed statistic for descriptor $k$ is the mean directional $R^2$ over seeds:
\[
T_{k,\mathrm{ang},\mathrm{obs}} =
\frac{1}{M}
\sum_{m=1}^{M}
R_{k,m}^2.
\]

This directional analysis should be interpreted as a global first-order summary. It may not capture nonlinear, radial, or multi-cluster organization of a descriptor.

\paragraph{Permutation tests.}

For statistical assessment, we use permutation tests that keep the learned embedding fixed and repeatedly break the association between embedding positions and the scalar values being tested. The permutation scheme depends on the dataset structure and on the analysis setting.

For the Sagar dataset, the $N=480$ subject--odorant observations are not fully independent, because some odorants are observed across multiple subjects. Standard analytic correlation p-values may therefore underestimate uncertainty. We instead employ restricted permutation tests, following the principle that permutations should preserve the dependence structure of the data \citep{winkler2014permutation,winkler2015multilevel}. The resulting null distribution describes how large the radial or directional statistic could be if the tested values were not systematically aligned with the embedding, while preserving selected aspects of the subject and odorant structure.

We propose two restricted permutation schemes for the Sagar union dataset, which test different null hypotheses. First, in the within-subject ($\mathrm{WS}$) permutation, values are shuffled separately for each subject. Given a subject $s$, let
\[
\mathcal{I}_s = \{o : (o,s)\in\mathcal{I}\}
\]
be the set of odorants observed for subject $s$. For each permutation $b=1,\ldots,B$, we draw a random permutation $\sigma_s^{(b)}$ of $\mathcal{I}_s$ and define
\[
y_{\left(o,s\right)}^{\left(b,\mathrm{WS}\right)}
=
y_{\left(\sigma_s^{\left(b\right)}\left(o\right),s\right)}.
\]
This permutation preserves each subject's distribution of values, but breaks the match between values and embedding positions within that subject. The corresponding null hypothesis is that, within each subject, the scalar values are exchangeable across odorants and are not specifically aligned with the embedding. Thus, the within-subject permutation tests whether the observed effect holds within subjects rather than being driven only by subject-specific rating biases or offsets.

Second, in the odorant-block ($\mathrm{OB}$) permutation, we shuffle whole odorant profiles. For each odorant $o$, we define its observed subject pattern as
\[
P(o) = \{s : (o,s)\in\mathcal{I}\}.
\]
For each subject pattern $P$, let
\[
\mathcal{O}_P = \{o : P(o)=P\}
\]
be the set of odorants with the same observed subject pattern. In the Sagar union dataset, the observed patterns are $\{1,2,3\}$, $\{1\}$, and $\{2,3\}$. For each pattern $P$, we draw a random permutation $\sigma_P^{(b)}$ of $\mathcal{O}_P$ and define
\[
y_{\left(o,s\right)}^{\left(b,\mathrm{OB}\right)}
=
y_{\left(\sigma_{P(o)}^{(b)}(o),s\right)}.
\]
This permutation preserves the subject profile of each odorant and the missingness structure of the repeated-measures data, but randomizes which odorant profile is attached to which embedding location. Intuitively, it is like swapping whole odorant ratings between compatible odorants: a rating observed for subjects $\{1,2,3\}$ can only be swapped with another rating observed for subjects $\{1,2,3\}$, and similarly for the other subject patterns. The corresponding null hypothesis is that odorant-level profiles are exchangeable among odorants with the same observed subject pattern and are not systematically aligned with the embedding.

For the GSLF dataset, each observation corresponds to one molecule and each molecule appears once. Therefore, there is no subject-level or repeated-measures structure to preserve. We use a molecule-level permutation test, denoted $p_{\mathrm{mol}}$. For each permutation $b=1,\ldots,B$, we draw a random permutation $\sigma^{(b)}$ of the molecule indices and define
\[
y_i^{(b,\mathrm{mol})}
=
y_{\sigma^{(b)}(i)}.
\]
The learned embedding is kept fixed, and the scalar values are shuffled across molecules. This tests whether the observed radial association is stronger than expected if the scalar values were exchangeable across molecules.

For radial analyses, the same procedure is applied to any scalar variable $y$. In Sagar, $y$ can be entropy or a continuous descriptor rating. In GSLF, $y$ is an entropy variable derived from the binary descriptor label. For each permutation scheme $Q$, where $Q$ can be $\mathrm{WS}$, $\mathrm{OB}$, or $\mathrm{mol}$ depending on the analysis, we recompute the mean Pearson radius--value correlation over seeds and denote the resulting permuted statistic by $T_{\mathrm{rad},b}^Q(y)$. The radial permutation p-value is two-sided because the correlation is signed:
\[
p_{\mathrm{rad},Q}(y)
=
\frac{1+N_{\mathrm{rad},Q}(y)}{B+1},
\]
where $N_{\mathrm{rad},Q}(y)$ is the number of permutations such that
\[
|T_{\mathrm{rad},b}^Q(y)|
\geq
|T_{\mathrm{rad},\mathrm{obs}}(y)|.
\]
A small radial p-value therefore indicates that the observed radial association is stronger than expected after breaking the association between scalar values and embedding radius under the corresponding permutation null model.

For the angular descriptor analysis in Sagar, the $\mathrm{WS}$ and $\mathrm{OB}$ permutation schemes are applied to the descriptor values $x_{\left(o,s\right),k}$. For each descriptor $k$ and each scheme $Q$, we refit the tangent-space linear regression after permutation and denote the resulting mean permuted $R^2$ over seeds by $T_{k,\mathrm{ang},b}^Q$. Since larger values of $R^2$ indicate stronger directional organization, the angular permutation p-value is one-sided:
\[
p_{k,\mathrm{ang},Q}
=
\frac{1+N_{k,\mathrm{ang},Q}}{B+1},
\]
where $N_{k,\mathrm{ang},Q}$ is the number of permutations such that
\[
T_{k,\mathrm{ang},b}^Q
\geq
T_{k,\mathrm{ang},\mathrm{obs}}.
\]
A small angular p-value indicates that the descriptor is better explained by a global tangent-space direction than expected after the descriptor values are permuted under the corresponding restricted null model.

In the Sagar union results, we report mainly $p_{\mathrm{WS}}$ and $p_{\mathrm{OB}}$. In the GSLF radial entropy results, we report $p_{\mathrm{mol}}$. The $+1$ correction avoids zero p-values when using a finite number of random permutations \citep{phipson2010permutation}.

\begin{table*}[h!]
\centering
\small
\begin{tabular}{lcccccc}
\toprule
Variable & Distance Pearson & Distance Spearman & Radial Pearson & Radial Spearman & $p_{\mathrm{WS}}$ & $p_{\mathrm{OB}}$ \\
\midrule
\emph{Rating profile entropy}
& \multirow{2}{*}{$0.82 \pm 0.02$}
& \multirow{2}{*}{$0.81 \pm 0.02$}
& $-0.77 \pm 0.05$
& $-0.72 \pm 0.05$
& $<0.001$
& $<0.001$ \\

Intensity descriptor rating
&
&
& $0.42 \pm 0.05$
& $0.38 \pm 0.05$
& $<0.001$
& $<0.001$ \\
\bottomrule
\end{tabular}
\caption{Main radial organization results for the Sagar union embedding. Values are reported as mean and standard deviation over 10 random seeds. Restricted permutation p-values were computed with $1000$ random permutations.}
\label{tab:sagar_main_result}
\end{table*}

\paragraph{Angular organization of binary descriptors: high-density regions.}

For a dataset with binary descriptor annotations, such as the GoodScents--Leffingwell dataset, descriptor values do not represent graded perceptual intensities, but only the presence or absence of a given odor label. In this setting, the tangent-space regression used above for continuous descriptors cannot be applied: a binary variable can be spatially concentrated, but it does not define a perceptual gradient as for a continuous rating. We therefore analyze binary descriptors through their spatial concentration in the embedding. Intuitively, if a binary odor label such as \emph{fruity} or \emph{floral} is meaningfully organized in the learned representation, then the molecules annotated with this label should occupy a coherent region of the disk rather than being uniformly scattered across the embedding. With this analysis, we evaluate whether different odor descriptors or descriptor families occupy coherent and distinguishable regions of the Poincaré disk. In particular, we ask whether the angular component of the learned hyperbolic representation reflects categorical structure in olfactory perception, with different angular sectors corresponding to different perceptual qualities.

This visualization is inspired by the descriptor-region visualizations used in the Principal Odor Map work \citep{POM_PrincipalOdorMap_Lee2023}. However, since our embedding space is hyperbolic, we adapt the density estimation and contour construction to the geometry of the Poincaré disk. Given a learned embedding \(\{z_i\}_{i=1}^{N}\), with \(z_i \in \mathbb{P}^2\), and a binary descriptor \(k\), let
\[
b_{i,k} \in \{0,1\}
\]
denote whether molecule \(i\) is annotated with descriptor \(k\). We define the set of molecules annotated with this descriptor as
\[
\mathcal{I}_k =
\{ i : b_{i,k}=1 \}.
\]
The goal is to estimate where, in the Poincaré disk, the molecules annotated with descriptor \(k\) are concentrated. We define a descriptor density by placing a smooth kernel around each positive example and averaging these kernels. For \(z \in \mathbb{P}^2\), this gives
\[
\widehat{\rho}_k(z)
=
\frac{1}{|\mathcal{I}_k|}
\sum_{i\in \mathcal{I}_k}
\exp\left(
-\frac{
d_{\mathbb{P}}\!\left(z,z_i\right)^2
}{
2h^2
}
\right),
\]
where \(h>0\) is a bandwidth hyperparameter controlling the smoothness of the density estimate. A larger value of \(h\) gives smoother and broader descriptor regions, whereas a smaller value gives sharper and more localized regions.

We then display the region where this descriptor density is highest. For a density threshold \(\lambda\), we define the corresponding upper-level region as
\[
\Omega_k(\lambda)
=
\left\{
z \in \mathbb{P}^2 :
\widehat{\rho}_k(z)
\geq
\lambda
\right\}.
\]
Intuitively, \(\Omega_k(\lambda)\) contains the points of the Poincaré disk where descriptor \(k\) has density at least \(\lambda\). Increasing \(\lambda\) keeps only the densest core of the descriptor, while decreasing \(\lambda\) gives a larger region.

The parameter we choose is not \(\lambda\) directly, but a target mass level
\[
\tau \in (0,1).
\]
For example, \(\tau=0.8\) means that we want to show the densest region containing approximately \(80\%\) of the estimated descriptor mass. The threshold \(\lambda_k(\tau)\) is then calculated as the density level whose upper-level region contains approximately a fraction \(\tau\) of the total descriptor mass.

Since the embedding lies in the Poincaré disk, this mass is computed using the hyperbolic area element \(dA_{\mathbb{P}}\). Thus, \(\lambda_k(\tau)\) is defined by
\[
\frac{
\int_{\Omega_k(\lambda_k(\tau))}
\widehat{\rho}_k(z)\,dA_{\mathbb{P}}(z)
}{
\int_{\mathbb{P}^2}
\widehat{\rho}_k(z)\,dA_{\mathbb{P}}(z)
}
\approx
\tau.
\]
The displayed descriptor region is therefore
\[
\Omega_k(\tau)
:=
\Omega_k(\lambda_k(\tau)).
\]
In other words, \(\Omega_k(\tau)\) is the high-density region of the Poincaré disk containing approximately a fraction \(\tau\) of the descriptor-specific hyperbolic KDE mass.

This construction allows us to visualize whether binary odor descriptors form coherent high-density regions in the learned hyperbolic representation, and whether different olfactory categories are associated with distinct angular sectors of the Poincaré disk.

\section{Results}

\subsection{Sagar}

The Sagar results are presented in three stages. We first report
the main radial and directional organization in the union
embedding, then evaluate robustness to alternative descriptor
representations, and finally examine subject-specific and
subject-averaged patterns.

\subsubsection{Main results}

\paragraph{Radial organization.}

\begin{figure*}[t]
\centering

\begin{minipage}[t]{0.77\textwidth}
\vspace{0pt}

    \begin{subfigure}[t]{0.49\linewidth}
        \centering
        \includegraphics[width=\linewidth]{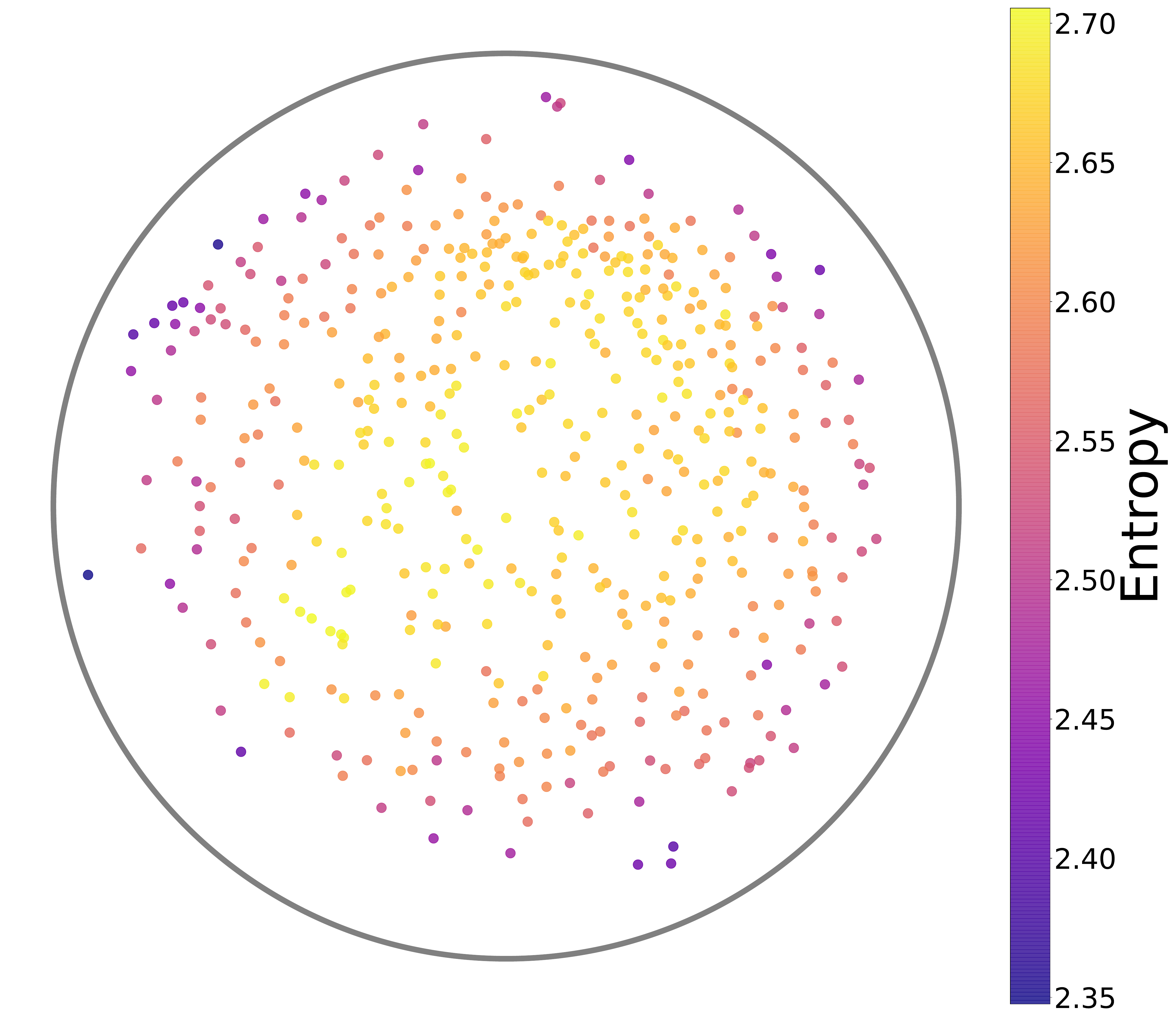}
        \caption{Embedding colored by \emph{rating profile entropy}}
        \label{fig:image1}
    \end{subfigure}
    \hfill
    \begin{subfigure}[t]{0.49\linewidth}
        \centering
        \includegraphics[width=\linewidth]{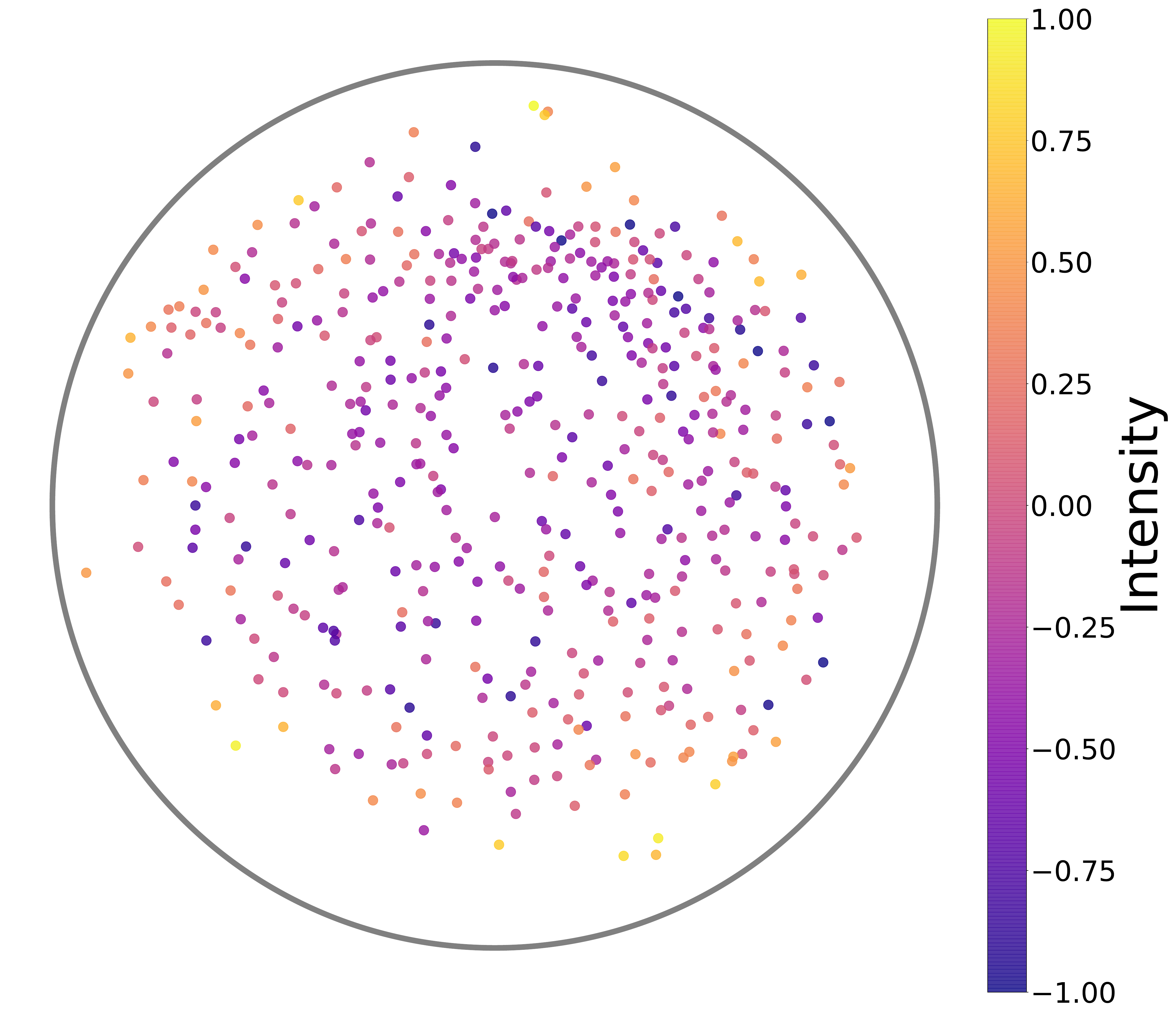}
        \caption{Embedding colored by intensity rating}
        \label{fig:image3}
    \end{subfigure}

    \vspace{0.5em}

    \begin{subfigure}[t]{0.45\linewidth}
        \centering
        \includegraphics[width=\linewidth]{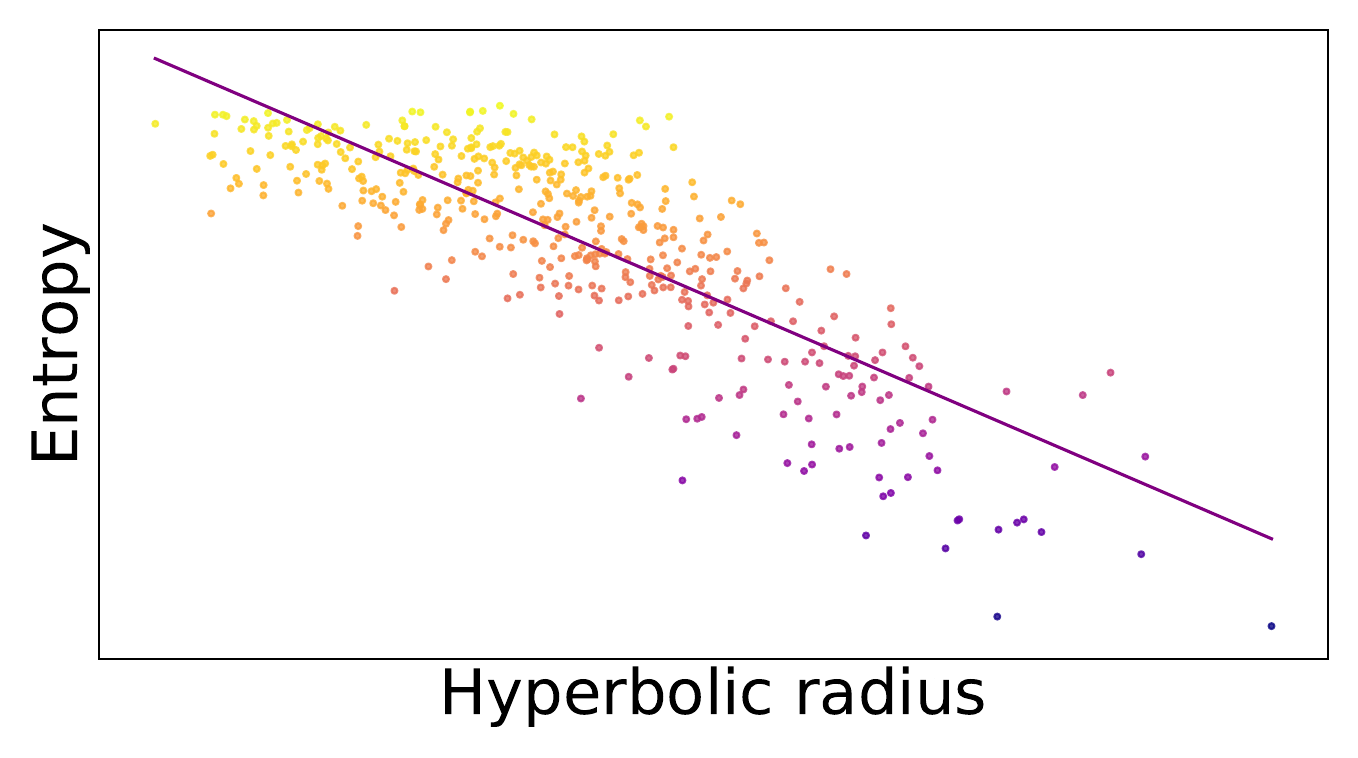}
        \caption{Entropy--radius correlation plot, Pearson correlation is -0.82 and Spearman is -0.78.}
        \label{fig:image2}
    \end{subfigure}
    \hfill
    \begin{subfigure}[t]{0.45\linewidth}
        \centering
        \includegraphics[width=\linewidth]{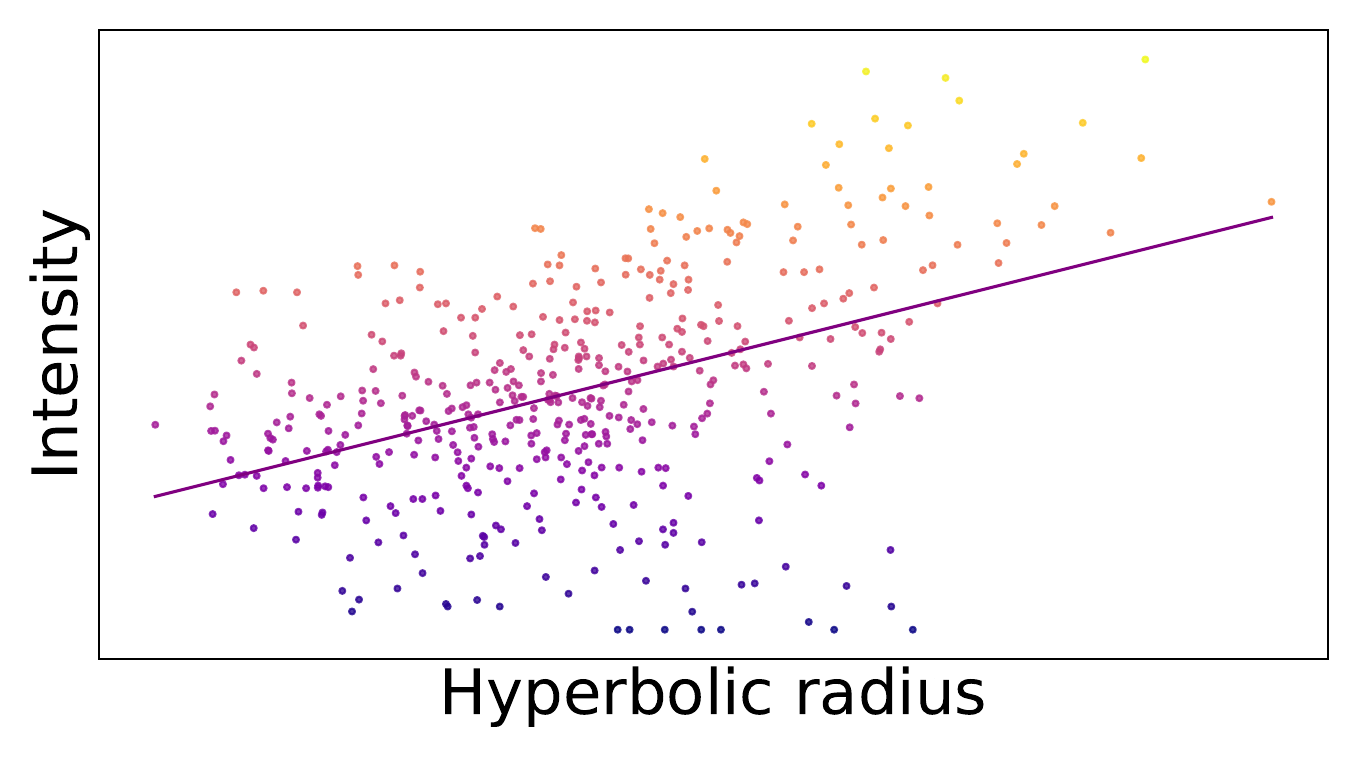}
        \caption{Intensity--radius correlation plot, Pearson correlation is 0.44 and Spearman is 0.41.}
        \label{fig:image4}
    \end{subfigure}

\end{minipage}
\hfill
%
\begin{minipage}[t]{0.2\textwidth}
\vspace{0pt}

\caption{
Representative \emph{Sagar} union embedding for random seed $m=5$.
Points are colored by \emph{rating profile entropy} and by intensity descriptor rating. Correlation plots show the corresponding radius associations.\\
\textbf{Alt text:} Four panels. \emph{Rating profile entropy} is higher near the center of the Poincaré disk and lower near the boundary, producing a strong negative correlation with radius. Intensity shows the opposite but weaker radial tendency.
}
\label{fig:radial_organization_Sagar_plots}

\end{minipage}

\end{figure*}

For the main configuration, trained on the original descriptor vectors, the learned hyperbolic embeddings preserved the input perceptual geometry well (Table~\ref{tab:sagar_main_result}). Distance Pearson and Spearman correlations compare pairwise distances in the input descriptor space with pairwise hyperbolic distances in the learned embedding. These embedding-level metrics are shared by the entropy and intensity analyses, and reached $0.82 \pm 0.02$ and $0.81 \pm 0.02$, respectively.

\emph{Rating profile entropy} computed from the original descriptors was strongly organized along the radial coordinate of the Poincaré disk. The radius--entropy correlation was strongly negative for both Pearson correlation, $-0.77 \pm 0.05$, and Spearman correlation, $-0.72 \pm 0.05$. This indicates that high-entropy observations tend to lie closer to the center, whereas low-entropy observations tend to lie closer to the boundary. This radial organization was supported by both restricted permutation tests, with $p_{\mathrm{WS}}<0.001$ and $p_{\mathrm{OB}}<0.001$. Figure~\ref{fig:radial_organization_Sagar_plots} illustrates these relationships for a representative random seed.

Among individual descriptors, intensity was the only descriptor showing a radial association (Table~\ref{tab:sagar_main_result}), with moderate radius--intensity correlations of $0.42 \pm 0.05$ for Pearson and $0.38 \pm 0.05$ for Spearman. The positive correlation indicates that higher-intensity observations tend to lie closer to the boundary. All other descriptors had radial correlations below $0.2$ in absolute value and are therefore not presented in the main table; the full descriptor-level results are reported in the Supplementary material (Table~\ref{tab:appendix_sagar_radial_all}). Importantly, the absolute radius--entropy correlation was substantially larger than the radius--intensity correlation, suggesting that the radial organization of the embedding is better explained by the \emph{rating profile entropy} over the full descriptor vector than by any single perceptual descriptor.

\paragraph{Angular organization of continuous descriptors.}

In addition to the radial organization of entropy and intensity, several descriptors showed angular organization in the learned hyperbolic embedding (Table~\ref{tab:sagar_angular_descriptors}). Directional $R^2$ quantifies how well each descriptor is explained by a global linear direction in the tangent-space representation of the Poincaré disk. Only descriptors with mean directional $R^2>0.2$ are shown in Table~\ref{tab:sagar_angular_descriptors}; the full descriptor-level table is reported in the Supplementary material (Table~\ref{tab:appendix_sagar_angular_all}).

The strongest directional trends were observed for \emph{sweet}, \emph{musky}, \emph{fruity}, and \emph{pleasantness}, with mean directional $R^2$ values ranging from $0.40$ to $0.51$. As visualized for \emph{sweet} and \emph{musky} in Figure \ref{fig:angular_organization_Sagar_plots} (additional figures for all other descriptors are available in the Supplementary material in Figure \ref{fig:all_descriptor_directions}), this indicates that these four descriptors vary primarily along dominant directions in the Poincaré disk rather than along the radial coordinate alone. All reported angular trends in Table \ref{tab:sagar_angular_descriptors} were supported by both restricted permutation tests, with $p_{\mathrm{WS}}<0.001$ and $p_{\mathrm{OB}}<0.001$.

\begin{figure}[h!]
    \centering

    \begin{subfigure}[t]{0.37\textwidth}
        \centering
        \includegraphics[width=\linewidth]{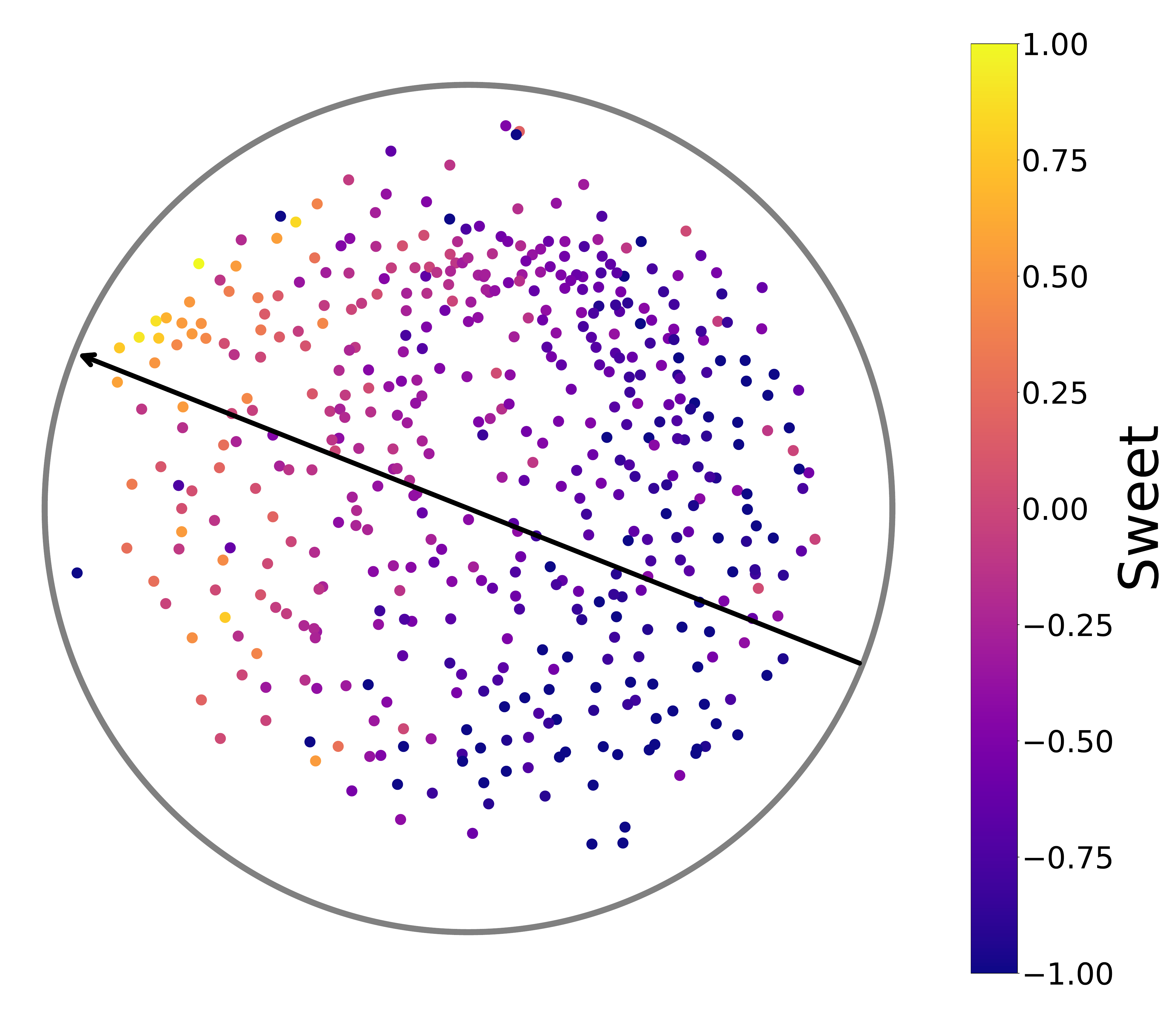}
        \caption{\emph{Sagar} embedding colored by sweet rating; directional angle at 158.40 degrees and $R^{2}=0.58$.}
        \label{fig:angular_sweet}
    \end{subfigure}
    \hfill
    \begin{subfigure}[t]{0.37\textwidth}
        \centering
        \includegraphics[width=\linewidth]{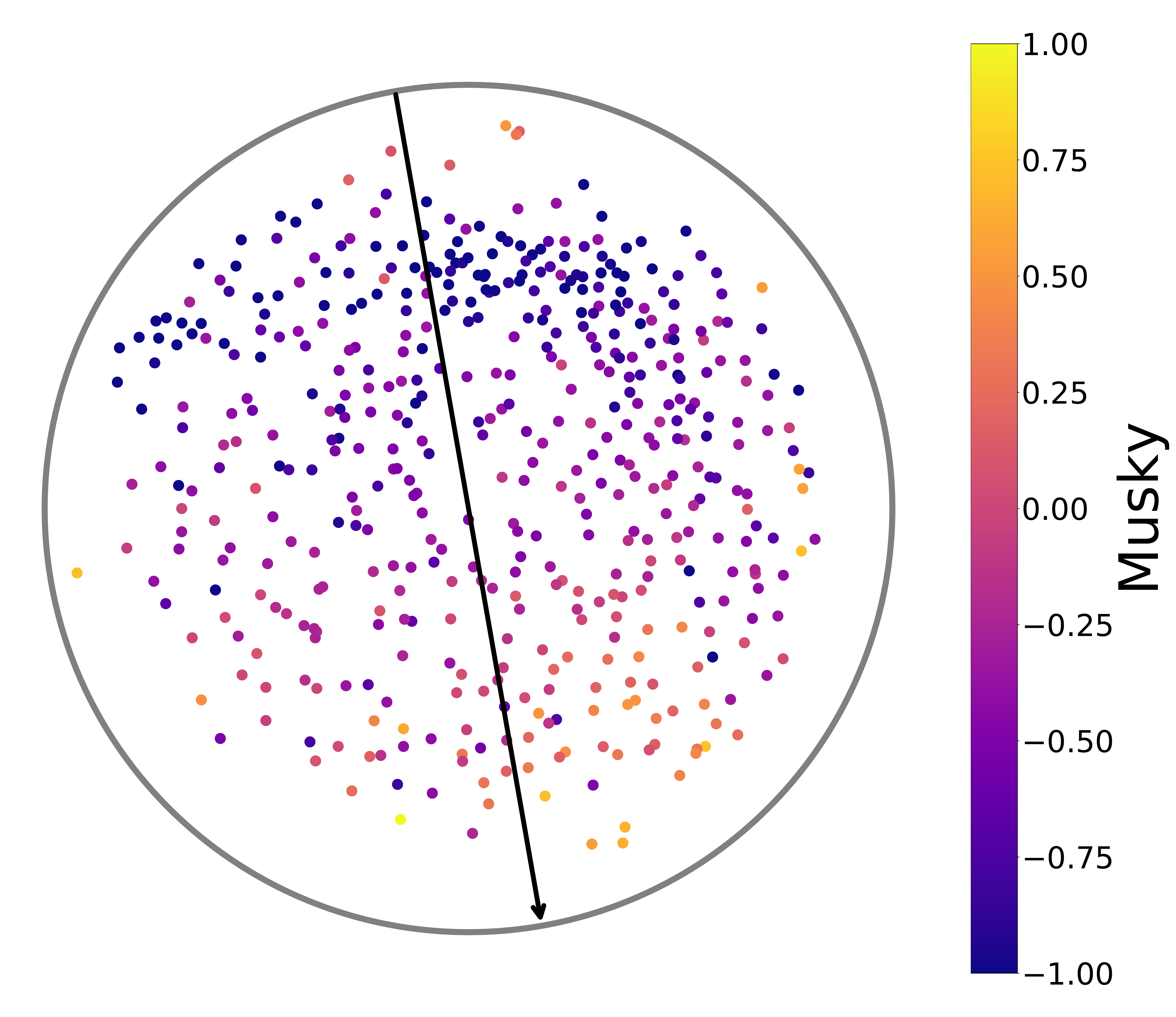}
        \caption{\emph{Sagar} embedding colored by musky rating; directional angle at 279.97 degrees and $R^{2}=0.43$.}
        \label{fig:angular_musky}
    \end{subfigure}
    \caption{Representative Sagar union embedding for random seed $m=5$, colored by \emph{sweet} and \emph{musky} ratings. Arrows indicate the fitted tangent space directions of steepest descriptor increase. The displayed angles and $R^2$ coefficient of determination values are seed specific.\\
    \textbf{Alt text:} Two Poincaré disk embeddings. Sweet ratings increase toward the upper left along a diagonal direction, whereas musky ratings increase downward along an approximately vertical direction. Both descriptors show clear directional organization.}
\label{fig:angular_organization_Sagar_plots}
\end{figure}

\begin{table}[h!]
\centering
\small
\begin{tabular}{lccc}
\toprule
Descriptor & Directional $R^2$ & $p_{\mathrm{WS}}$ & $p_{\mathrm{OB}}$ \\
\midrule
Sweet & $0.51 \pm 0.12$ & $<0.001$ & $<0.001$ \\
Musky & $0.51 \pm 0.06$ & $<0.001$ & $<0.001$ \\
Fruity & $0.46 \pm 0.09$ & $<0.001$ & $<0.001$ \\
Pleasantness & $0.40 \pm 0.09$ & $<0.001$ & $<0.001$ \\
Decayed & $0.34 \pm 0.06$ & $<0.001$ & $<0.001$ \\
Warm & $0.26 \pm 0.12$ & $<0.001$ & $<0.001$ \\
Floral & $0.25 \pm 0.11$ & $<0.001$ & $<0.001$ \\
Bakery & $0.25 \pm 0.13$ & $<0.001$ & $<0.001$ \\
Fishy & $0.23 \pm 0.04$ & $<0.001$ & $<0.001$ \\
\bottomrule
\end{tabular}
\caption{Angular organization results for the Sagar union embedding, sorted by decreasing mean directional $R^2$. Values are reported as mean and standard deviation over 10 random seeds. Restricted permutation p-values were computed with $1000$ random permutations.}
\label{tab:sagar_angular_descriptors}
\end{table}

\subsubsection{Ablation and robustness analyses of the radial entropy organization}

\begin{table*}[t]
\centering
\scriptsize
\setlength{\tabcolsep}{10pt}
\renewcommand{\arraystretch}{1}
\begin{tabular}{@{}llcccccc@{}}
\toprule
Training input & Entropy type & Dist. Pearson & Dist. Spearman & Radial Pearson & Radial Spearman & $p_{\mathrm{WS}}$ & $p_{\mathrm{OB}}$ \\
\midrule
\multirow{3}{*}{Original descriptors}
& Descriptors excl. intensity
& \multirow{3}{*}{$0.82 \pm 0.02$}
& \multirow{3}{*}{$0.81 \pm 0.02$}
& $-0.75 \pm 0.06$
& $-0.70 \pm 0.06$
& $<0.001$
& $<0.001$ \\

& Pruned descriptors
&
&
& $-0.68 \pm 0.04$
& $-0.65 \pm 0.05$
& $<0.001$
& $<0.001$ \\

& Orthogonalized descriptors
&
&
& $-0.69 \pm 0.07$
& $-0.79 \pm 0.09$
& $<0.001$
& $<0.001$ \\
\midrule

Descriptors excl. intensity
& Descriptors excl. intensity
& $0.83 \pm 0.02$
& $0.81 \pm 0.02$
& $-0.74 \pm 0.05$
& $-0.69 \pm 0.06$
& $<0.001$
& $<0.001$ \\

Pruned descriptors
& Pruned descriptors
& $0.82 \pm 0.01$
& $0.81 \pm 0.01$
& $-0.72 \pm 0.05$
& $-0.66 \pm 0.08$
& $<0.001$
& $<0.001$ \\
\bottomrule
\end{tabular}
\caption{Ablation and robustness analysis of the radius--entropy relationship in the Sagar union embedding. Values are reported as mean and standard deviation over 10 random seeds. For the first three rows, the same original-input embedding is used; therefore, the distance-preservation metrics are shared. Restricted permutation p-values were computed with $1000$ random permutations.}
\label{tab:sagar_entropy_ablation}
\end{table*}

To test whether the radius--entropy relationship depends on the particular descriptor representation, we performed an ablation and robustness analysis (Table~\ref{tab:sagar_entropy_ablation}). This analysis addresses three possible concerns. First, since intensity was the only individual descriptor with a non-negligible radial association, the entropy effect could potentially be driven by intensity. Second, the effect could depend on redundant or correlated descriptors in the original descriptor space. Third, the effect could depend on the original descriptor coordinate system.

We tested these possibilities in two complementary ways. In the first three rows of Table~\ref{tab:sagar_entropy_ablation}, the embedding is kept fixed and trained on the original descriptors, while the entropy variable is recomputed using alternative descriptor representations: descriptors excluding intensity, pruned descriptors, and orthogonalized descriptors. This isolates the effect of changing the entropy definition without changing the learned embedding geometry. The pruned descriptor representation was constructed to reduce descriptor redundancy: descriptors were removed greedily until no remaining pair of descriptors had absolute correlation higher than $0.3$. The removed descriptors were pleasantness, decayed, musky, fruity, sweet, and bakery. In the last two rows, the embedding itself is retrained using either descriptors excluding intensity or pruned descriptors, and entropy is computed in the corresponding reduced descriptor space. This provides a stronger test of whether the radius--entropy organization persists when the removed descriptors are excluded from the geometry used to learn the embedding.

The radius--entropy relationship remained strong and negative across all configurations. When intensity was removed only from the entropy computation, the radial Pearson correlation remained close to the main result, with $-0.75 \pm 0.06$, and the radial Spearman correlation was $-0.70 \pm 0.06$. When intensity was removed both from the training input and from the entropy computation, the correlations remained similarly strong, with radial Pearson correlation $-0.74 \pm 0.05$ and radial Spearman correlation $-0.69 \pm 0.06$. This indicates that the entropy-radius relationship is not merely an intensity effect.

The effect also persisted when entropy was computed on pruned descriptors, and when the embedding was retrained using the pruned descriptor space. In both cases, the radial correlations remained negative and substantial, with Pearson correlations between $-0.68$ and $-0.72$. Finally, entropy computed from orthogonalized descriptors also showed a strong radial relationship, with radial Pearson correlation $-0.69 \pm 0.07$ and radial Spearman correlation $-0.79 \pm 0.09$. Thus, the radial entropy organization is not specific to the original descriptor basis.

Across all ablation and robustness configurations, the distance-preservation metrics remained high, with distance Pearson and Spearman correlations around $0.81$--$0.83$. All radius--entropy associations were supported by both restricted permutation tests, with $p_{\mathrm{WS}}<0.001$ and $p_{\mathrm{OB}}<0.001$. Overall, these results support the interpretation that hyperbolic radius primarily reflects the entropy of the perceptual descriptor profile, rather than being driven by a single descriptor, by descriptor redundancy, or by the original coordinate system.

\subsubsection{Subject-level and averaged-rating robustness analyses}

We next tested whether the radial entropy organization and angular descriptor organization were stable across subjects, and whether they were also present at the level of subject averaged ratings. The above studied main Sagar union embedding contains subject--odorant observations from all three subjects. Therefore, an apparent radial entropy effect could in principle be influenced by subject-level pooling effects, for example if one subject systematically produced higher-entropy ratings and was also placed closer to the center of the embedding.

We considered two complementary settings. First, we analyzed each subject separately within the union embedding. In this case, no new embedding was trained: for each random seed, we used the embedding trained on the full union dataset and then restricted the analysis to the $160$ observations belonging to a single subject. This tests whether the entropy-radius relationship is visible within each subject's observations in the shared union geometry. Second, we trained new embeddings on averaged descriptor ratings. For this averaged-rating analysis, descriptor vectors were averaged across subjects for each CID, retaining the $125$ CIDs observed in all three subjects. This gives one consensus descriptor vector per odorant. We then trained hyperbolic embeddings on these averaged descriptor vectors and computed entropy from the averaged descriptor profile. This tests whether the radial and angular organization is also present at the level of averaged-ratings odor perception, after removing subject-specific rating variability.

For these robustness analyses, we used the same permutation logic as in the main evaluation, but adapted the exchangeability unit to the setting considered. For subject-level analyses within the union embedding, we used a within-subject permutation, denoted $p_{\mathrm{WS}}$, by shuffling the tested variable across odorants within the selected subject. For the averaged-rating embedding, each point corresponds to one CID, so we used a CID-level permutation, denoted $p_{\mathrm{CID}}$, by shuffling the tested variable across CIDs. Here, the tested variable is entropy for the radial analysis and the descriptor value for the angular analysis. Radial entropy p-values test whether the radius--entropy correlation is unusually strong in either direction, whereas angular p-values test whether the directional $R^2$ is unusually large.

\begin{figure*}[t]
\centering

\begin{subfigure}[t]{0.245\textwidth}
    \centering
    \includegraphics[width=\linewidth]{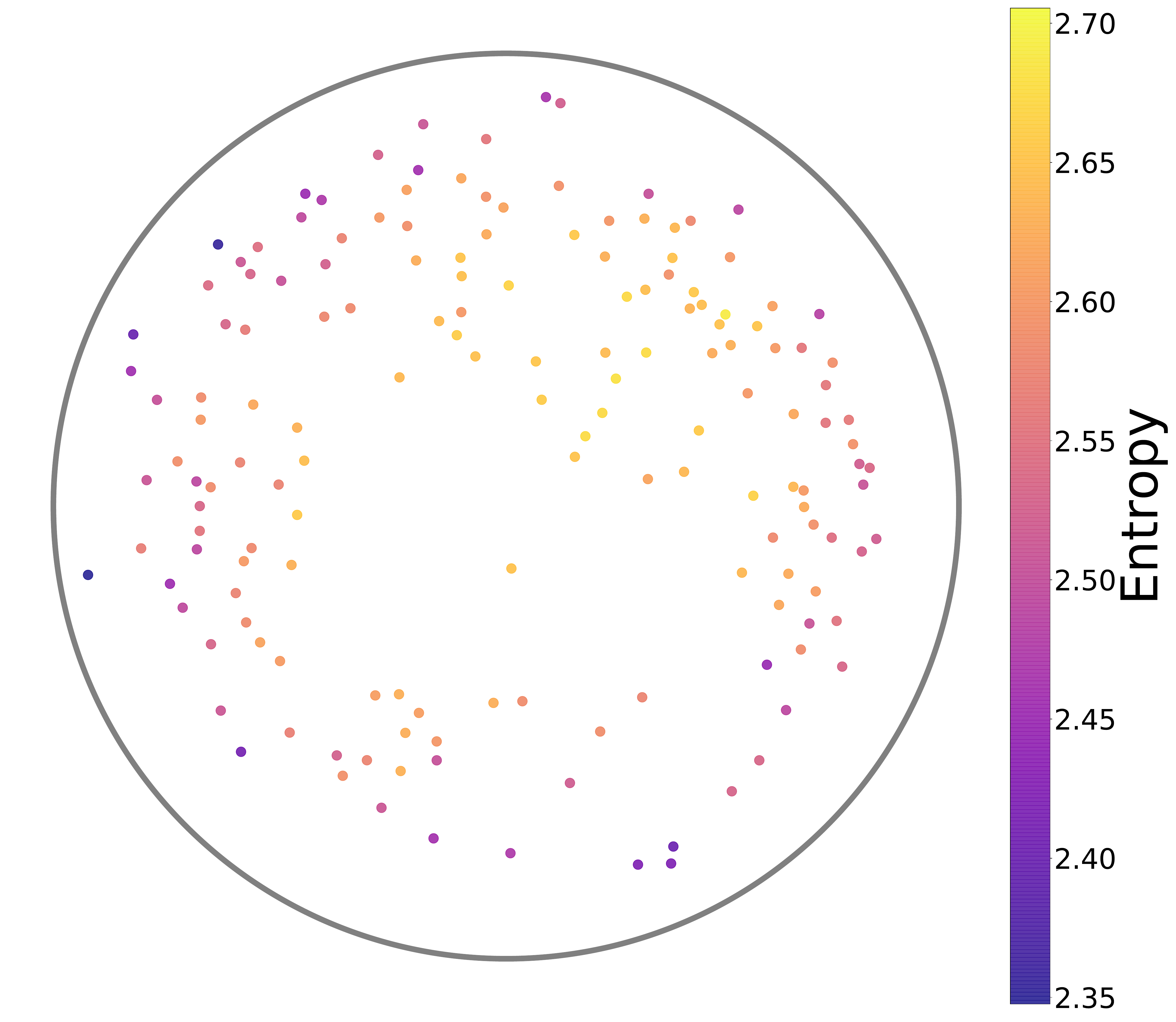}
    \caption{Subject 1. Pearson correlation is -0.78 and Spearman is -0.78.}
    \label{fig:subj1}
\end{subfigure}
\hfill
\begin{subfigure}[t]{0.245\textwidth}
    \centering
    \includegraphics[width=\linewidth]{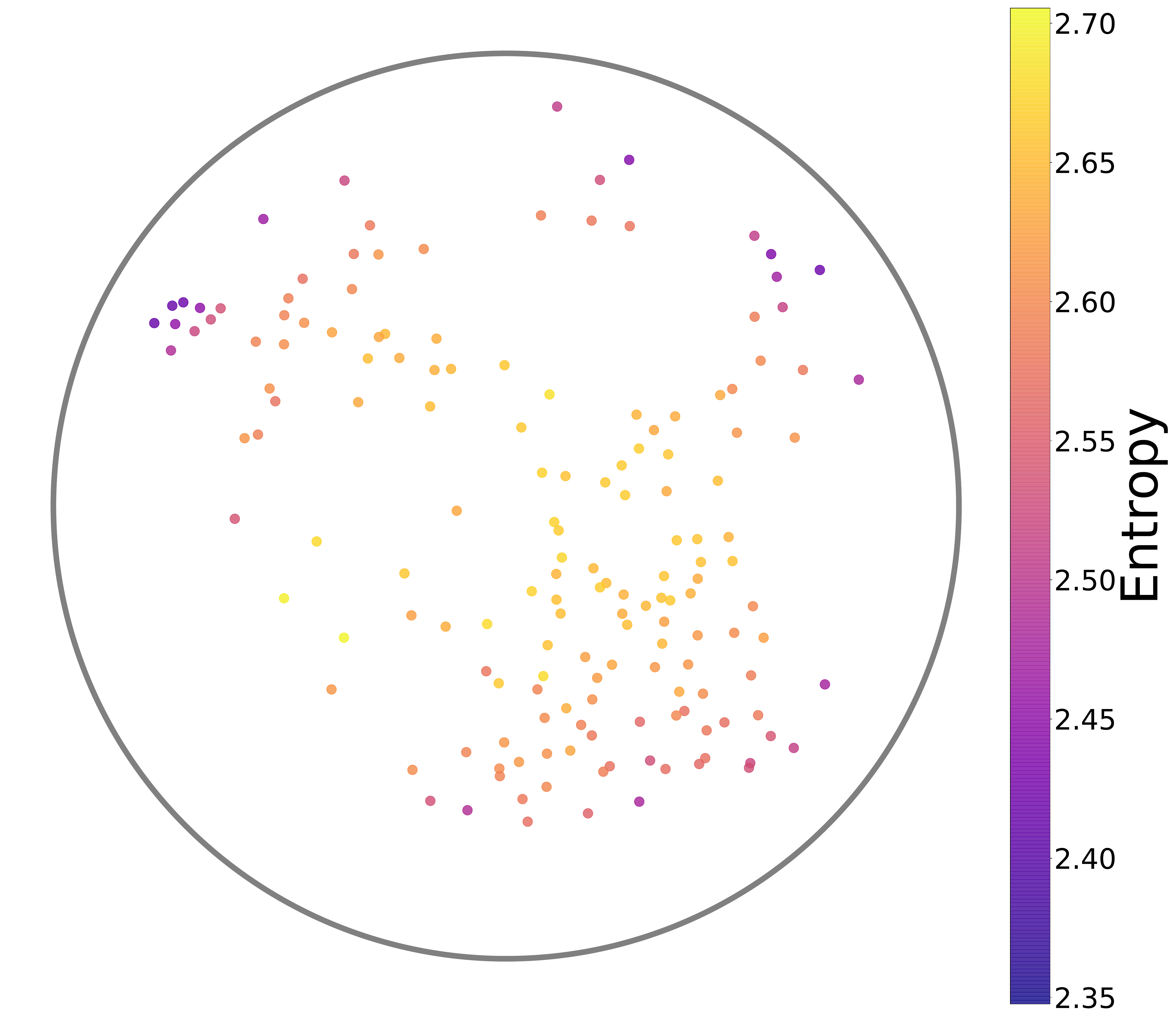}
    \caption{Subject 2. Pearson correlation is -0.90 and Spearman is -0.90.}
    \label{fig:subj2}
\end{subfigure}
\hfill
\begin{subfigure}[t]{0.245\textwidth}
    \centering
    \includegraphics[width=\linewidth]{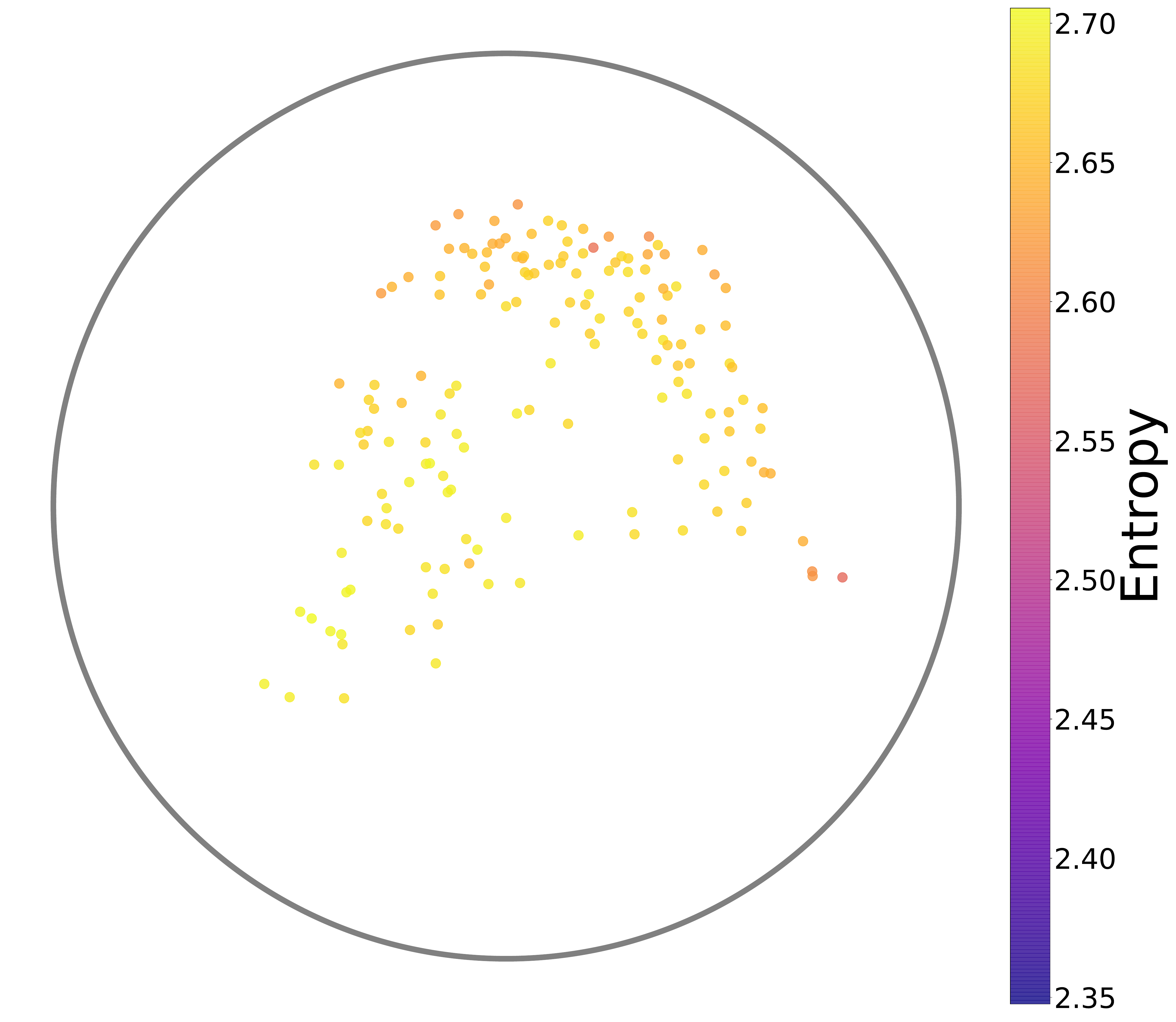}
    \caption{Subject 3. Pearson correlation is -0.65 and Spearman is -0.66.}
    \label{fig:subj3}
\end{subfigure}
\hfill
\begin{subfigure}[t]{0.245\textwidth}
    \centering
    \includegraphics[width=\linewidth]{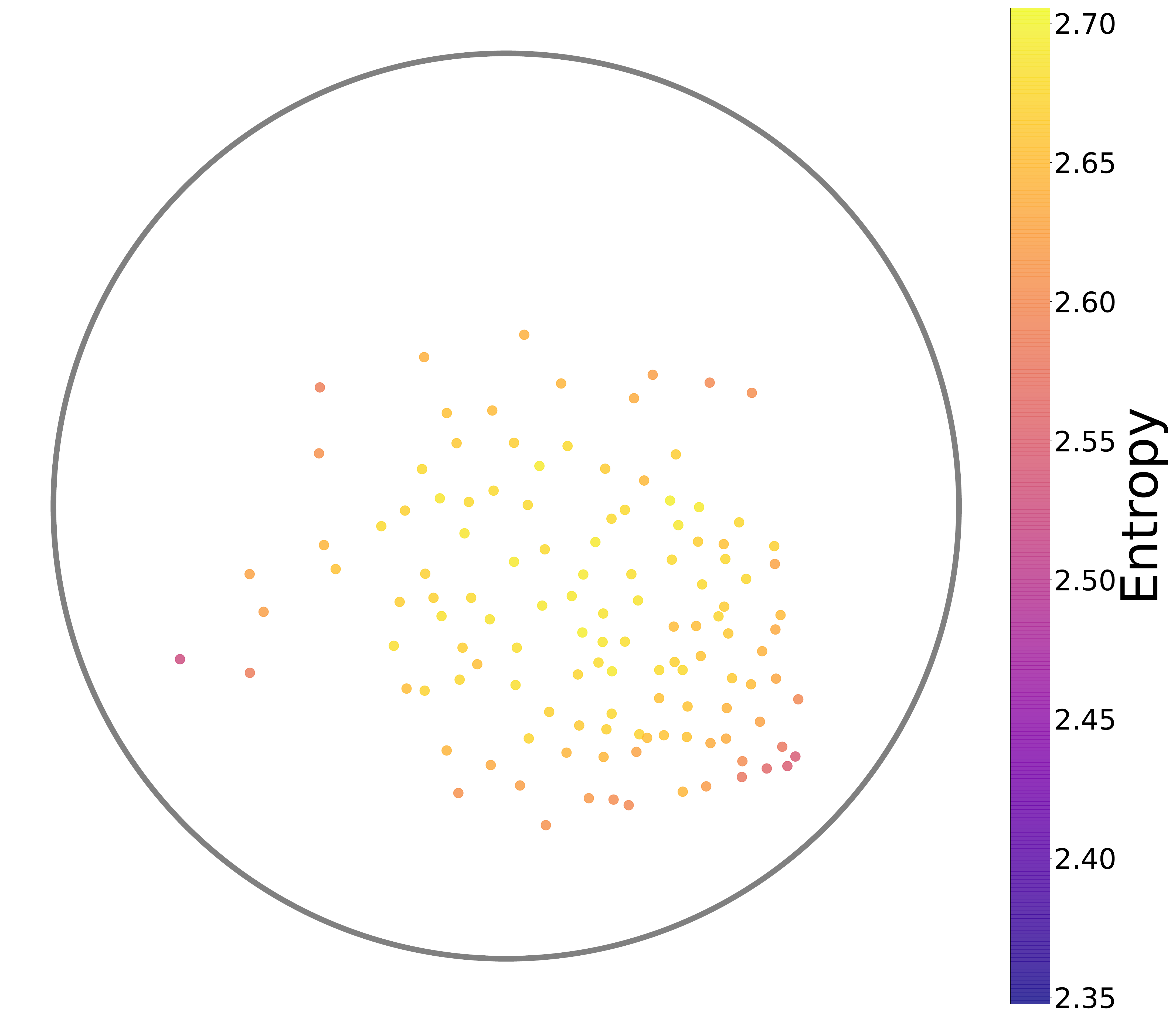}
    \caption{Averaged ratings. Pearson is -0.77 and Spearman is -0.72.}
    \label{fig:average}
\end{subfigure}

\caption{Poincaré embeddings colored by \emph{rating profile entropy} for the \emph{Sagar} dataset (random seed $m=5$): individual subjects and averaged ratings.\\
\textbf{Alt text:} Four embeddings for the three individual subjects and averaged ratings. In every case, higher rating profile entropy occurs predominantly toward the center and lower entropy toward the boundary, although the spatial distributions differ.}
\label{fig:radial_organization_Sagar_plots_subjects_and_average}

\end{figure*}

\paragraph{Radial entropy organization.}
Table~\ref{tab:sagar_subject_average_radial_entropy} reports the subject-level and averaged-rating radial profile entropy results. When the union embedding was analyzed separately within each subject, the entropy-radius relationship remained strongly negative: the radial Pearson correlations were $-0.72 \pm 0.07$ for subject 1, $-0.86 \pm 0.03$ for subject 2, and $-0.61 \pm 0.06$ for subject 3. This shows that the radial entropy organization is not merely a consequence of pooling subjects in the union dataset, but is already present within each subject's observations.

\begin{table}[h]
\centering
\small
\scriptsize
\setlength{\tabcolsep}{9pt}
\renewcommand{\arraystretch}{1}
\begin{tabular}{lcccc}
\toprule
Analysis & $N_{\mathrm{CID}}$ & Radial Pearson & Radial Spearman & $p$ \\
\midrule
Subject 1 & $160$ & $-0.72 \pm 0.07$ & $-0.69 \pm 0.10$ & $<0.001$ \\
Subject 2 & $160$ & $-0.86 \pm 0.03$ & $-0.87 \pm 0.03$ & $<0.001$ \\
Subject 3 & $160$ & $-0.61 \pm 0.06$ & $-0.63 \pm 0.09$ & $<0.001$ \\
Average & $125$ & $-0.82 \pm 0.08$ & $-0.78 \pm 0.12$ & $<0.001$ \\
\bottomrule
\end{tabular}
\caption{Subject-level and averaged-rating robustness analysis of the radial entropy organization. Values are reported as mean and standard deviation over 10 random seeds. The p-value column reports $p_{\mathrm{WS}}$ for subject-level rows and $p_{\mathrm{CID}}$ for the averaged-rating row, both computed with $1000$ random permutations.}
\label{tab:sagar_subject_average_radial_entropy}
\end{table}

The strength of this organization nevertheless varied across individuals. Subject 2 showed the strongest radius--entropy association, whereas subjects 1 and 3 showed weaker but still substantial associations. This variability may reflect differences in rating behavior, perceptual strategy, or the structure of each subject's odor perceptual space. Importantly, the averaged-rating embedding also showed a strong negative radius--entropy correlation, $-0.82 \pm 0.08$, indicating that the effect is preserved at the level of averaged-ratings odor perception despite individual variability. This radial--entropy relationship, across individual subjects and at the averaged level, is illustrated in Figure \ref{fig:radial_organization_Sagar_plots_subjects_and_average} which shows the different embeddings for one random seed.

\begin{figure}[h!]
\centering
\includegraphics[width=1\linewidth]{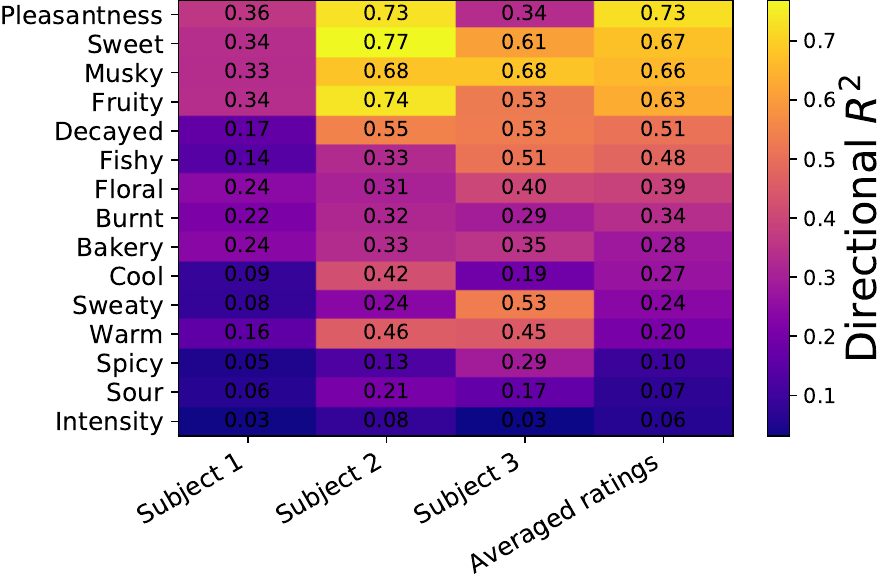}
\caption{Subject-level and averaged-rating angular descriptor organization. Each entry of this heatmap shows the mean directional $R^2$ over 10 random seeds.\\
\textbf{Alt text:} Heatmap comparing directional $R^{2}$ across descriptors for three subjects and averaged ratings. 
}
\label{fig:subject_average_angular_R2_heatmap}
\end{figure}

\paragraph{Angular organization of continuous descriptors.}
Figure~\ref{fig:subject_average_angular_R2_heatmap} summarizes the corresponding angular descriptor organization. The first three columns show the union embedding restricted to one subject at a time, whereas the last column shows embeddings trained directly on descriptor ratings averaged across subjects. Directional $R^2$ quantifies how well each descriptor is explained by a global tangent-space direction in the Poincaré disk.

Several descriptor directions were stable across each of the three available subjects and in the averaged-rating embedding. Pleasantness, sweet, musky, fruity, floral, burnt, and bakery all had mean directional $R^2>0.2$ in all four configurations: subject 1, subject 2, subject 3, and the averaged-rating embedding. These seven descriptors also had permutation p-values below $0.001$ in all four configurations. The strongest and most consistent angular directions were associated with pleasantness, sweet, musky, and fruity, in agreement with the union-level angular analysis reported in Table~\ref{tab:sagar_angular_descriptors}. This supports the interpretation that angular position captures meaningful perceptual descriptor gradients, while radial position is primarily associated with the entropy of the descriptor profile.

The angular organization also revealed subject-level variability. Subject 2 generally showed stronger directional organization than subjects 1 and 3, especially for pleasantness, sweet, fruity, and musky. Subject 3 showed strong angular organization for fishy, decayed, musky, fruity, sweaty, sweet, and warm, whereas subject 1 showed weaker but still detectable directional trends. Thus, while similar descriptor families tend to define angular directions across subjects, the strength of these directions differs between individuals. This suggests that angular coordinates capture perceptual descriptor gradients that are partly shared across subjects, but also modulated by individual rating patterns.

\begin{table*}[h!]
\centering
\small
\begin{tabular}{lccccc}
\toprule
Entropy type & Dist. Pearson & Dist. Spearman & Radial Pearson & Radial Spearman & $p_{\mathrm{mol}}$ \\
\midrule
\emph{Active label entropy}
& \multirow{3}{*}{$0.69 \pm 0.01$}
& \multirow{3}{*}{$0.68 \pm 0.01$}
& $0.87 \pm 0.03$
& $0.88 \pm 0.03$
& $<0.001$ \\

Pruned \emph{active label entropy}
&
&
& $0.78 \pm 0.03$
& $0.78 \pm 0.03$
& $<0.001$ \\

\emph{Orthogonalized descriptor entropy}
&
&
& $-0.88 \pm 0.01$
& $-0.92 \pm 0.02$
& $<0.001$ \\
\bottomrule
\end{tabular}
\caption{Radial entropy organization results for the \emph{GSLF} dataset. Distance Pearson and Spearman correlations are shared embedding quality metrics because computed from the same learned embeddings. Values are reported as mean and standard deviation over 10 random seeds. Molecule level permutation p-values, denoted $p_{\mathrm{mol}}$, were computed with $1000$ random permutations.}
\label{tab:gslf_radial_entropy}
\end{table*}

\begin{figure}[h!]
    \centering

    \begin{subfigure}[t]{0.30\textwidth}
        \centering
        \includegraphics[width=\linewidth]{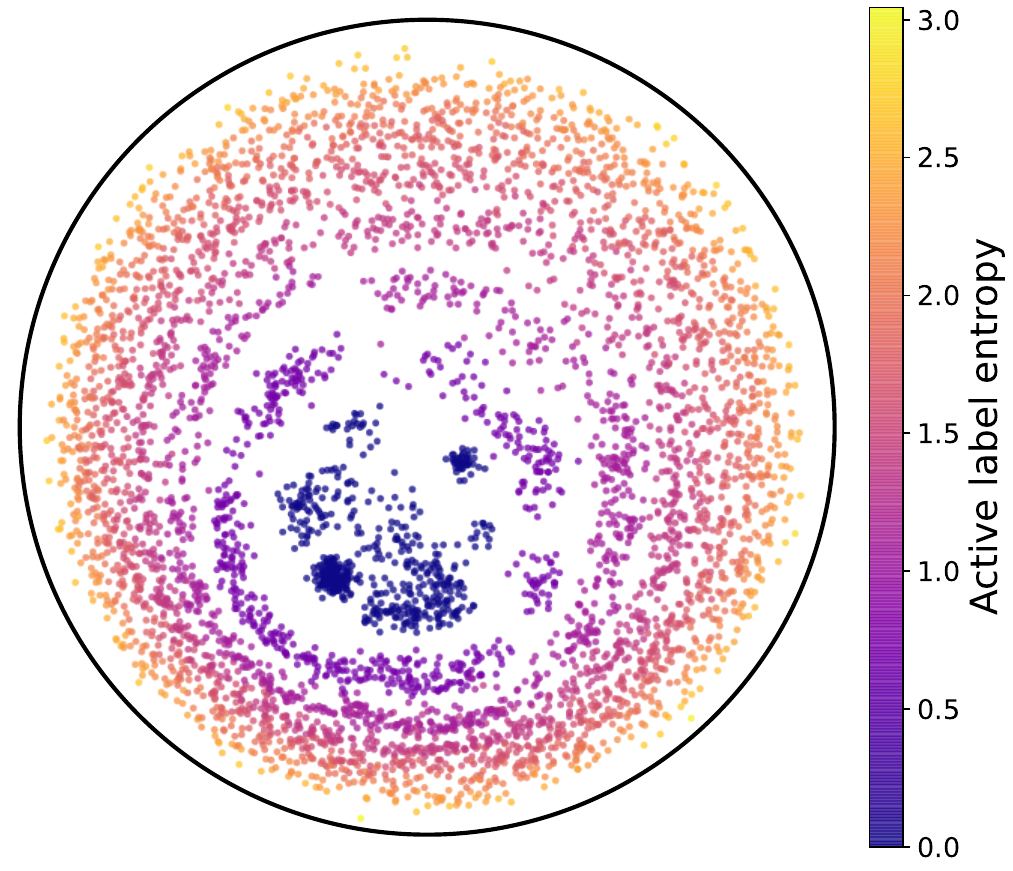}
        \caption{Embedding colored by \emph{active label entropy}}
        \label{fig:imagegslf1}
    \end{subfigure}
    \hfill
    \begin{subfigure}[t]{0.29\textwidth}
        \centering
        \includegraphics[width=\linewidth]{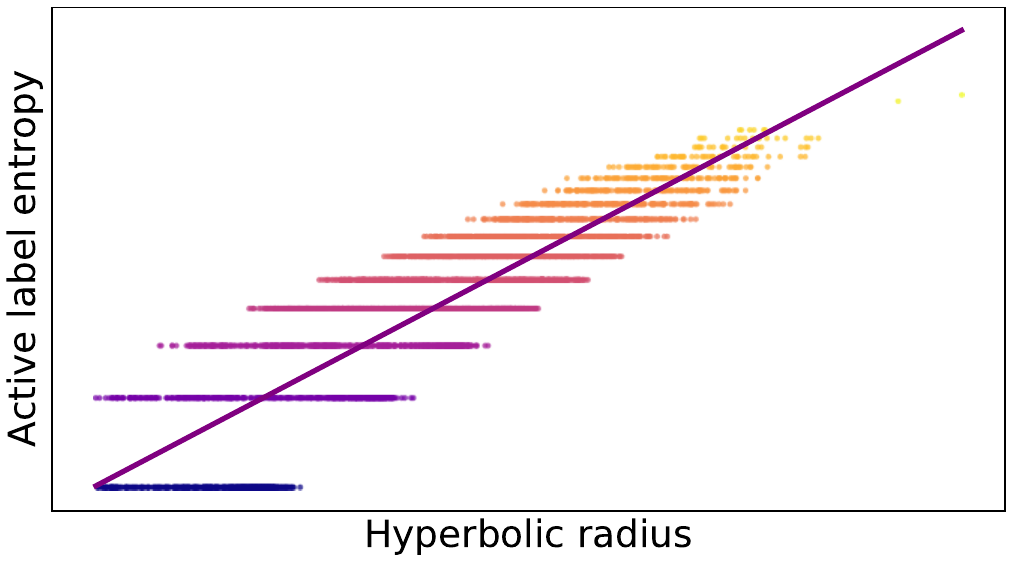}
        \caption{Radius correlation plot for \emph{active label entropy}, Pearson is 0.86 and Spearman is 0.88.}
        \label{fig:imagegslf2}
    \end{subfigure}
    \hfill
    \begin{subfigure}[t]{0.30\textwidth}
        \centering
        \includegraphics[width=\linewidth]{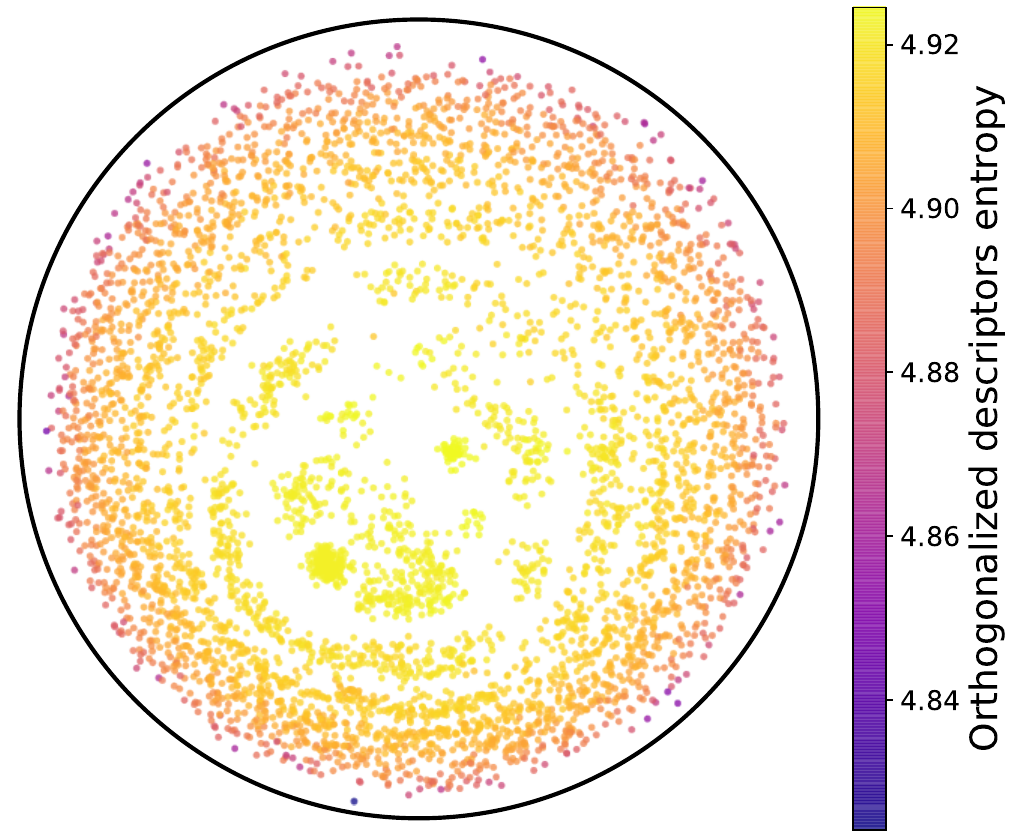}
        \caption{Embedding colored by \emph{orthogonalized descriptor entropy}}
        \label{fig:imagegslf3}
    \end{subfigure}
    \hfill
    \begin{subfigure}[t]{0.29\textwidth}
        \centering
        \includegraphics[width=\linewidth]{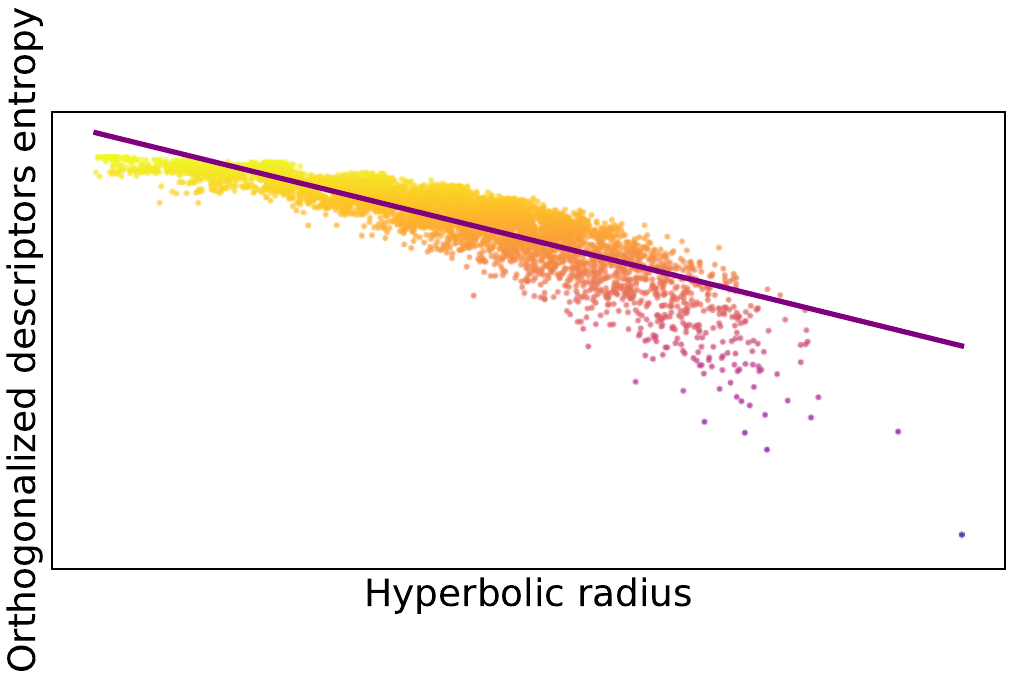}
        \caption{Radius correlation plot for \emph{orthogonalized descriptor entropy}, Pearson is -0.87 and Spearman is -0.91.}
        \label{fig:imagegslf4}
    \end{subfigure}

    \caption{
    Representative GSLF embedding and radius associations for \emph{active label entropy} and \emph{orthogonalized descriptor entropy}, at random seed $m=1$.\\
    \textbf{Alt text:} Four panels showing radial entropy patterns in the GSLF embedding. \emph{Active label entropy} increases strongly from the center toward the boundary, whereas \emph{orthogonalized descriptor entropy} decreases strongly with radius.}
    \label{fig:radial_organization_gslf_plots}
\end{figure}

\subsection{GoodScents--Leffingwell (GSLF)}

The GSLF dataset provides a complementary test of the proposed geometric organization in a larger expert annotated odor dataset. Unlike Sagar, GSLF descriptors are binary multi-label annotations rather than continuous subject ratings. Therefore, entropy has a different interpretation. \emph{Active label entropy} measures the breadth of the binary descriptor profile, whereas \emph{orthogonalized descriptor entropy} measures whether the transformed descriptor profile is diffuse across several orthogonal modes of variation or dominated by one or a few modes.

\paragraph{Radial entropy organization.}

Table~\ref{tab:gslf_radial_entropy} reports the radial entropy results for the GSLF embedding. As an embedding quality check, the learned hyperbolic representations preserved the input binary descriptor geometry with distance Pearson correlation $0.69 \pm 0.01$ and distance Spearman correlation $0.68 \pm 0.01$. These values indicate that the embedding retains a substantial part of the pairwise descriptor structure, although the distance preservation is lower than in the Sagar dataset. This could be explained by the sparse binary nature of the GSLF descriptor matrix and its higher dimensionality (138 descriptors for GSLF compared to 15 for Sagar).

\emph{Active label entropy} showed a strong positive association with hyperbolic radius, with radial Pearson correlation $0.87 \pm 0.03$ and radial Spearman correlation $0.88 \pm 0.03$. This indicates that molecules annotated with broader descriptor profiles tend to lie closer to the boundary of the Poincaré disk. Importantly, this result should not be interpreted in the same way as the negative radius--entropy correlation observed in Sagar. In Sagar, \emph{rating profile entropy} measures the spread of continuous descriptor strengths, whereas in GSLF \emph{active label entropy} measures the number of active binary labels. Thus, the positive correlation in GSLF suggests that molecules associated with many odor qualities occupy more peripheral regions of the embedding.

The radial organization remained strong after reducing descriptor redundancy. When \emph{active label entropy} was computed from the pruned descriptor representation, where remaining labels have absolute correlation inferior or equal to 0.3, the radial Pearson and Spearman correlations remained high with both at $0.78 \pm 0.03$. This indicates that the association between \emph{active label entropy} and radius is not solely driven by correlated or redundant descriptors.

Finally, \emph{orthogonalized descriptor entropy} showed a strong negative association with hyperbolic radius, with radial Pearson correlation $-0.88 \pm 0.01$ and radial Spearman correlation $-0.92 \pm 0.02$. This result has a different interpretation from \emph{active label entropy}. After orthogonalization, the dimensions no longer correspond to individual odor labels, but to orthogonal modes of variation in the descriptor data. \emph{Orthogonalized descriptor entropy} therefore tests whether the transformed score profile is balanced across several modes or dominated by one or a few modes. This negative correlation is consistent with the Sagar results when entropy is computed from continuous or orthogonalized descriptor profiles. It indicates that, in the orthogonalized descriptor representation, molecules closer to the center have more diffuse profiles across orthogonal modes, whereas molecules closer to the boundary have profiles dominated by one or a few modes.

All three radial entropy associations were significant under molecule-level permutation testing, with $p_{\mathrm{mol}}<0.001$. Overall, these results show that the radial coordinate of the GSLF embedding captures entropy-related structure, but the interpretation depends on the entropy definition. \emph{Active label entropy} reflects descriptor multiplicity, whereas \emph{orthogonalized descriptor entropy} reflects spread across orthogonal modes of variation. Figure \ref{fig:radial_organization_gslf_plots} illustrates these relationships for a representative random seed.

\paragraph{Angular organization of binary descriptors.}

We next examined whether binary odor descriptors occupy coherent regions of the GSLF embedding. Because GSLF labels are binary annotations, they do not define graded descriptor directions in the same sense as the continuous ratings in Sagar. We therefore used the hyperbolic KDE visualization described in the Method section to identify high density regions associated with descriptor families.

\begin{figure}[h!]
\centering
\includegraphics[width=0.95\linewidth]{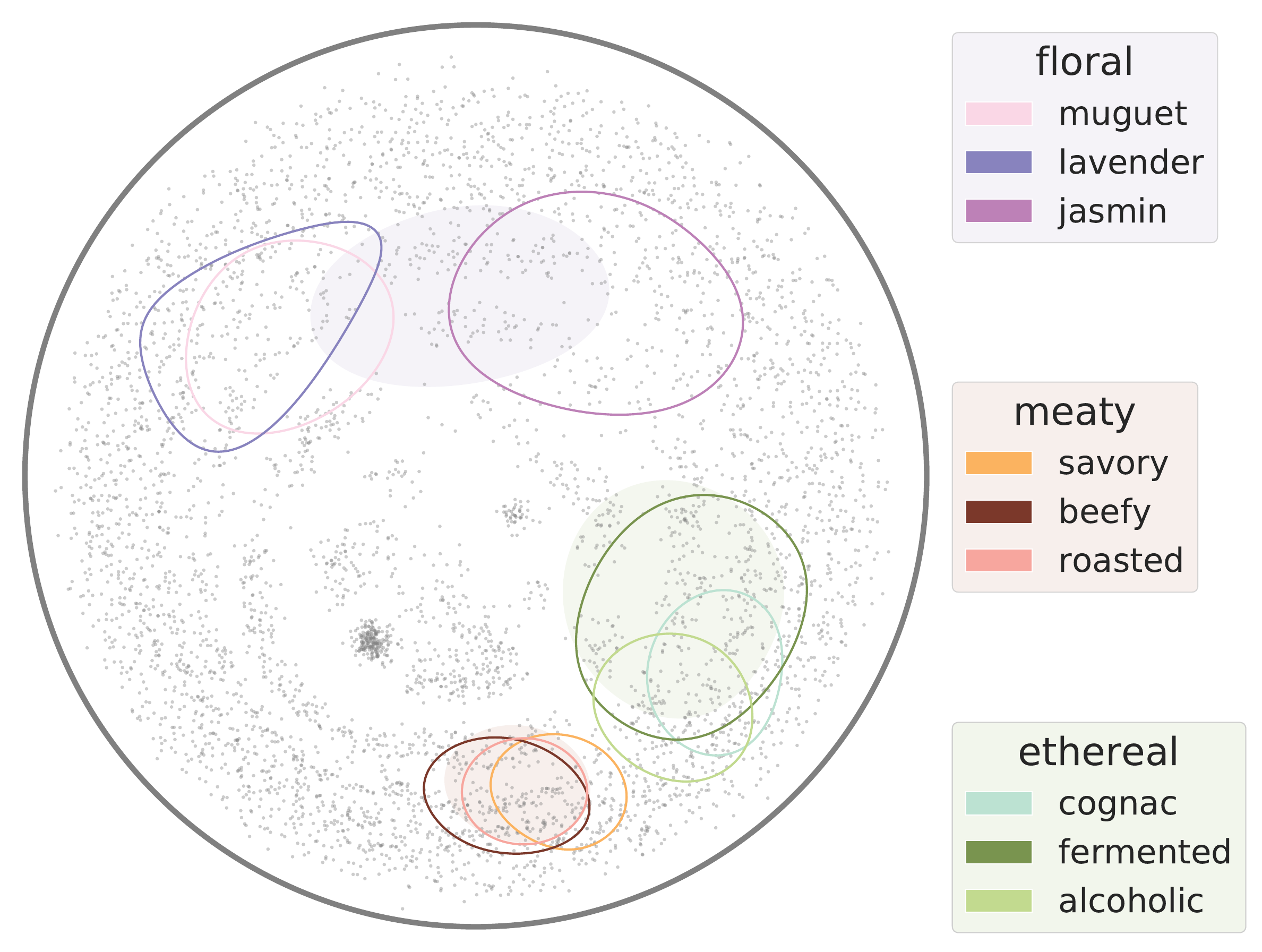}
\caption{High density regions for representative GSLF descriptor families in the Poincaré disk. Gray points show all molecules in the embedding. Filled regions and contours show hyperbolic KDE upper level sets for broad descriptor families and related subdescriptors, using a KDE mass level of $\tau=5\%$.\\
\textbf{Alt text:} Poincaré disk showing spatially separated descriptor families. \emph{Floral} descriptors occupy the upper region, \emph{meaty} descriptors the lower region, and \emph{ethereal} descriptors the lower right region. Related subdescriptors form overlapping localized contours within each family.}
\label{fig:gslf_3classes}
\end{figure}

Following \cite{POM_PrincipalOdorMap_Lee2023}, Figure~\ref{fig:gslf_3classes} shows representative high density regions for three broad descriptor families: \emph{floral}, \emph{meaty}, and \emph{ethereal}. Molecules annotated with related descriptors tend to occupy nearby regions of the Poincaré disk. Floral descriptors such as \emph{floral}, \emph{muguet}, \emph{lavender}, and \emph{jasmin} form a coherent region, whereas meaty descriptors such as \emph{meaty}, \emph{savory}, \emph{beefy}, and \emph{roasted} occupy a distinct region. Ethereal descriptors such as \emph{ethereal}, \emph{cognac}, \emph{fermented}, and \emph{alcoholic} form a third region. These three descriptor families are spatially separated and occupy different angular sectors of the disk, suggesting that angular position captures categorical structure among binary odor labels.

\begin{figure}[h!]
\centering
\includegraphics[width=1\linewidth]{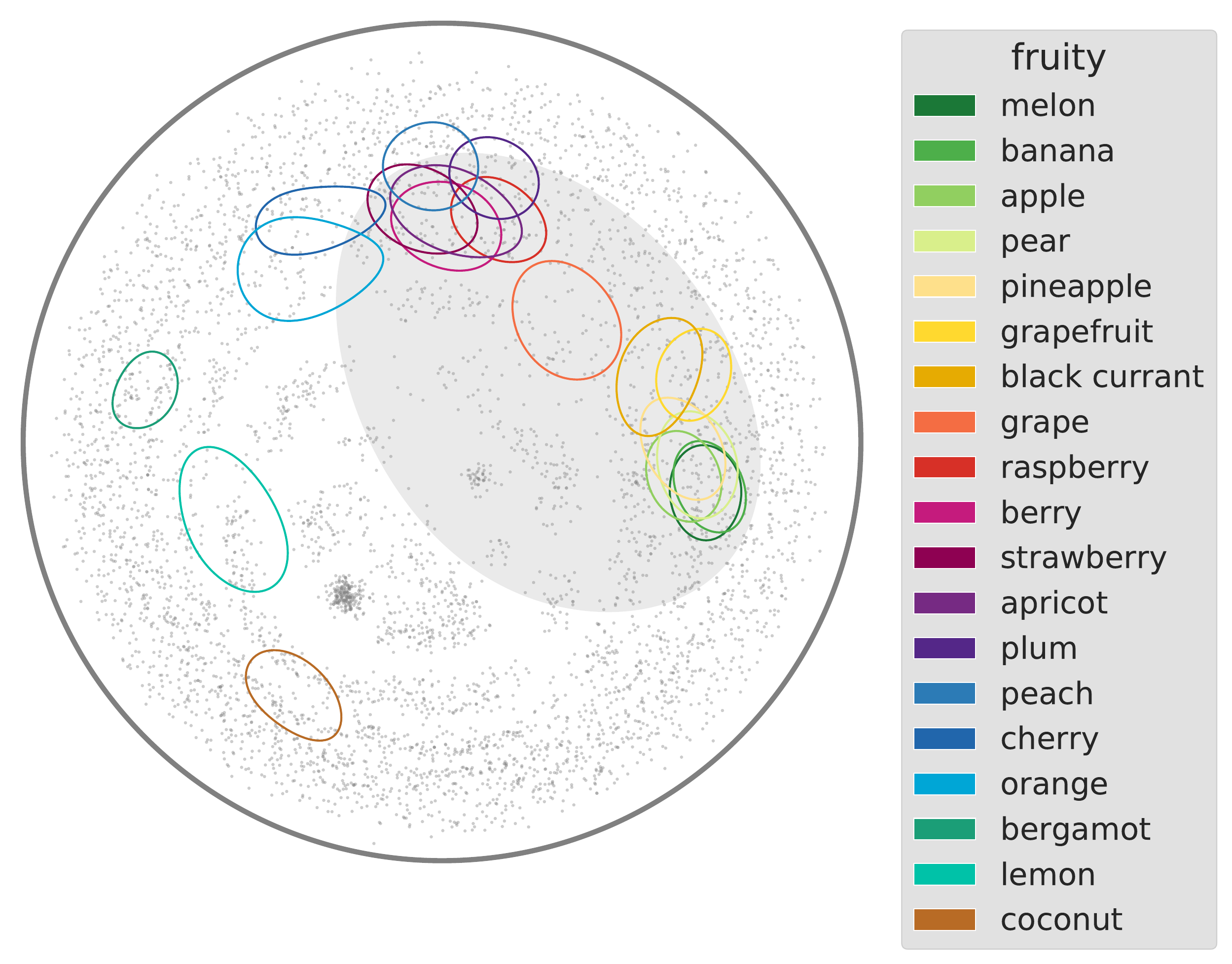}
\caption{High density regions for the \emph{fruity} descriptor family in the GSLF embedding. Gray points show all molecules. The filled region corresponds to the broader \emph{fruity} descriptor and is shown with KDE mass level $\tau=20\%$. Contours correspond to fruit related subdescriptors and are shown with KDE mass level $\tau=1\%$.\\
\textbf{Alt text:} Poincaré disk showing a broad \emph{fruity} descriptor region surrounded by localized fruit subdescriptor contours. Many fruit descriptors cluster within or near the main fruity region, while \emph{bergamot}, \emph{lemon}, and \emph{coconut} form more separated regions.}
\label{fig:gslf_fruits}
\end{figure}

Figure~\ref{fig:gslf_fruits} provides a finer grained visualization of the fruity descriptor family. The broader \emph{fruity} region contains or overlaps with several fruit related descriptor regions, including \emph{melon}, \emph{banana}, \emph{apple}, \emph{pear}, \emph{pineapple}, \emph{grapefruit}, \emph{black currant}, \emph{grape}, \emph{raspberry}, \emph{berry}, \emph{strawberry}, \emph{apricot}, \emph{plum}, \emph{peach}, \emph{cherry}, \emph{orange}. By contrast, \emph{bergamot}, \emph{lemon}, and \emph{coconut} appear farther from the main fruity region. Thus, the embedding captures both broad odor families and finer categorical distinctions within a family: related fruit descriptors tend to occupy a common sector of the disk, while individual descriptors remain locally distinguishable.

These visualizations suggest a complementary organization of the GSLF embedding. The radial coordinate is associated with entropy related properties of descriptor profiles, whereas angular position appears to reflect categorical olfactory structure, separating broad odor families and organizing finer subcategories within them.

\section{Discussion}

\subsection{Summary and interpretation}

The present work shows that olfactory descriptor data exhibit complementary radial and angular organization when represented
in the two-dimensional Poincaré disk. In the Sagar dataset, hyperbolic radius was most strongly associated with \emph{rating profile entropy}, whereas individual descriptors were better characterized by directional trends. The radial association persisted across the robustness analyses, supporting the interpretation that radius reflects a global property of the
descriptor profile rather than any single perceptual descriptor. More diffuse profiles were located closer to the center, whereas
more concentrated profiles were located closer to the boundary.

The directional analysis provided a complementary description of odor quality. Several descriptors, including \emph{sweet},
\emph{musky}, \emph{fruity}, \emph{pleasantness}, and \emph{decayed}, were well summarized by dominant tangent-space directions. In particular, the strong directional trend of \emph{pleasantness} is consistent with previous work identifying pleasantness as an important organizing axis of olfactory perception \citep{crocker1927analysis, Khan2007PredictingOdorPleasantness, koulakov2011search, snitz2013predicting, licon2018pleasantness}. The hyperbolic representation therefore retained an established perceptual dimension while revealing a distinct radial organization related to the descriptor profile as a whole. Similar directional patterns were observed in the subject-specific and subject-averaged analyses, although their strength varied among the three subjects.

The GSLF results extended this geometric decomposition to a larger dataset of binary expert annotations. In this setting, \emph{active label entropy} increased with radius, indicating that molecules assigned a larger number of odor descriptors tended to occupy more peripheral regions of the disk. This positive association does not contradict the negative radius--entropy relationship observed in Sagar, because
\emph{active label entropy} measures descriptor multiplicity, whereas \emph{rating profile entropy} measures how diffusely continuous rating strength is distributed across descriptors. A complementary pattern emerged within GSLF itself: \emph{orthogonalized descriptor entropy} decreased with radius, showing that molecules closer to the center had more diffuse profiles across orthogonal modes of variation, whereas those closer to the boundary were dominated by fewer modes, in line with Sagar. Thus, the direction of the radial association depends on the property summarized by the entropy measure, while radius consistently captures global structure in the descriptor profile. In addition, related binary descriptors occupied coherent high-density regions, with broad odor families and finer subcategories
appearing in distinct angular sectors of the disk.

Together, these findings support hyperbolic mapping as an interpretable descriptive framework in which radius summarizes global properties of descriptor profiles, while the angular component captures descriptor-specific gradients through directional trends in continuous ratings and categorical organization through localized high-density regions for binary descriptors.

\subsection{Limitations and future work}

The present study has several limitations that also point toward useful directions for future work. First, the present analysis is based on descriptor data and does not directly test neural mechanisms of olfactory coding. Although the results are compatible with the hypothesis that olfactory perception has non-Euclidean structure, linking these geometric features to neural representations will require analyses of brain data, such as fMRI or EEG, collected from a sufficiently large and diverse participant sample.

A further limitation is the small number of subjects in the continuous rating dataset. The subject specific analyses provide initial evidence that the observed radial and directional organization is not restricted to a single rating profile. However, larger and more diverse samples will be needed to characterize the consistency of these patterns and their variability across the broader population. Such studies could also benefit from richer descriptor vocabularies provided by trained assessors or odor experts. Indeed, hyperbolic spaces can be seen as continuous analogs of trees \citep{Krioukov2010_HyperbolicGeometryComplexNetworks} and thus, hyperbolic geometry may be most informative when descriptors are organized into an explicit multilevel taxonomy with substantial branching and depth, rather than as a flat list of broad odor qualities. Data of this kind would make it possible to test more directly whether hyperbolic embeddings capture hierarchical relations among odor categories, for example via hyperbolic tree geometric inference methods such as \cite{ MedbouhiGarcia2026RandomizedHyperSteiner}.

The spatial analysis of binary descriptors should also be interpreted cautiously. The hyperbolic KDE regions used for binary descriptors are qualitative visualizations. They provide descriptive evidence that odor labels occupy coherent regions of the disk, but they do not constitute formal statistical tests of category separation. Future work could complement these visualizations with quantitative measures of spatial concentration, overlap, and separation between descriptor families.

Beyond the perceptual descriptor space itself, the present study does not incorporate molecular structure. Integrating molecular features with perceptual descriptors in a joint hyperbolic framework could help determine how chemical similarity relates to the radial entropy organization and angular odor category structure observed here. Such a model could also clarify which aspects of the learned geometry arise from perceptual judgments and which are already present in the molecular organization of the odorants.

Finally, the present work does not include behavioral confidence ratings, emotion measures, cognitive style measures, personality measures, or clinical assessments of olfactory function. The differences observed in the subject specific analyses should therefore be interpreted as differences in rating patterns, rather than as evidence for particular cognitive, personality, or sensory mechanisms. Previous work nevertheless suggests that olfactory perception and confidence in sensory judgments may be influenced by personality traits \citep{shepherd2017personality, seo2013relationships}, including neuroticism \citep{croy2011agreeable}, agreeableness and openness to experience \citep{tyagi2024differences}, as well as by emotion and cognitive biases \citep{chen2005effect}. The strong radial association observed here indicates that entropy is a global organizing property of olfactory descriptor profiles in the learned perceptual representation. However, because entropy was computed from the ratings rather than directly judged by the participants, its psychological interpretation remains tentative. Studies with larger samples could combine descriptor ratings with direct judgments of perceptual clarity, complexity, ambiguity, and confidence to investigate whether unusually high entropy reflects uncertainty or less differentiated judgments, and whether unusually low entropy reflects confidence, overconfidence, or a restricted response strategy. Separately, psychophysical and clinical measures of olfactory function could be employed to examine whether embeddings confined to a limited region of the Poincaré disk are associated with reduced perceptual differentiation or olfactory impairment, rather than with differences in vocabulary, scale use, or rating strategy.

\section*{Conflicts of interest}
No relevant conflict of interest declared.

\section*{Funding}
This work has been supported by the Swedish Research Council, Knut and Alice Wallenberg Foundation, and the European Research Council (ERC-2023-SyG 10118977 D2Smell).

\section*{Acknowledgements}
The authors wish to thank Pawel Andrzej Herman for his valuable feedback. The authors acknowledge the use of artificial intelligence (AI) tools to assist with language editing and code development. All AI assisted code was reviewed, tested, and validated by the authors. The authors retained full responsibility for the scientific design, analyses, interpretations, and conclusions.

\section*{Data availability}
The Sagar dataset is publicly available through the Pyrfume repository \citep{hamel2024pyrfume}, and the GoodScents--Leffingwell dataset is available through OpenPOM \citep{OpenPOM}. Our code used to generate the embeddings, perform the geometric and statistical analyses, and reproduce the figures is available at \url{https://github.com/anissmedbouhi/HyperSmell}.

\bibliographystyle{ccn_style}

\bibliography{ccn_style}

\newpage

\clearpage
\onecolumn
\appendix

\section*{Supplementary material}

\setcounter{table}{0}
\setcounter{figure}{0}

\renewcommand{\thetable}{S\arabic{table}}
\renewcommand{\thefigure}{S\arabic{figure}}

\section{Hyperbolic geometry and optimization details}
\label{app:hyperbolic_details}

This Supplementary material provides the geometric expressions and
optimization details underlying the hyperbolic metric MDS model. The main text contains the concepts needed to understand the radial and directional analyses, whereas the general Poincaré ball expressions and Riemannian optimization updates
are provided here for reproducibility.

\subsection{Geometry of the Poincaré ball}
\label{app:poincare_geometry}

We start to describe the model where we embed our data, and explicit the metric tensor and the derived hyperbolic distance. Formally, the $n$-dimensional hyperbolic space is the unique simply-connected Riemannian manifold with constant curvature equal to $-1$. The hyperbolic space admits several models; in this work, we focus on the \emph{Poincaré ball} model. The latter is the Riemannian manifold given by the Euclidean ball
\[
\mathbb{P}^n = \left\{z \in \mathbb{R}^n \mid \|z\| < 1 \right\}, 
\]
where $\|\cdot\|$ denotes the Euclidean norm, equipped with a Riemannian metric consisting of the Euclidean inner product scaled by a factor that reflects the curvature of the space.

At a point $z\in\mathbb{P}^{n}$, we define the conformal factor as
\[
\lambda_z
=
\frac{2}{1-\|z\|^2}.
\]
For tangent vectors $v,w\in T_z\mathbb{P}^{n}$ identified with $\mathbb{R}^n$, the Riemannian metric is then
\[
g_z(v,w)
=
\lambda_z^2\langle v,w\rangle,
\]
where $\langle\cdot,\cdot\rangle$ denotes the ordinary Euclidean
inner product. The corresponding Riemannian norm of a tangent vector is
\[
\|v\|_z
=
\sqrt{g_z(v,v)}
=
\lambda_z\|v\|.
\]

As for any Riemannian manifold, $\mathbb{P}^n$ can be seen as a metric space when equipped with the geodesic distance, i.e., the length of the shortest path between two points of the manifold. Formally, the geodesic distance between two points $x,y\in\mathbb{P}^{n}$ is defined as: 
\[
d_{\mathbb{P}}(x,y) = \inf_\gamma \int_{[0,1]} \sqrt{g_{\gamma(t)}(\gamma'(t), \gamma'(t))} \ dt,
\]
where ${\gamma\colon [0,1] \rightarrow \mathbb{P}^n}$ is a smooth curve with ${\gamma(0) = x}$, ${\gamma(1)=y}$, and $\gamma'$ denotes the first derivative of $\gamma$. Explicitly, the distance can be computed via the simple expression:
\[
d_{\mathbb{P}}(x, y) = \operatorname{arcosh}\left(1 + \frac{2\|x - y\|^2}{\left(1 - \|x\|^2\right)\left(1 - \|y\|^2\right)}\right).
\]
where $\operatorname{arcosh}(r) = \ln \left( r + \sqrt{r^2 - 1} \right)$, for $r \geq 1$.

In particular, the hyperbolic radius of a point $z$ is
\[
d_{\mathbb{P}}(0,z)
=
2\operatorname{artanh}(\|z\|).
\]
Consequently, equal Euclidean displacements correspond to
increasingly large hyperbolic distances as points approach the
boundary of the disk.

\subsection{Möbius addition}
\label{app:mobius_addition}

The exponential map, logarithmic map, and parallel transport can
be written using Möbius addition. For $x,y\in\mathbb{P}^{n}$,
Möbius addition is defined as \citep{Ungar2009Gyrovector}:
\[
x\oplus y
=
\frac{
\left(
1+2\langle x,y\rangle+\|y\|^2
\right)x
+
\left(
1-\|x\|^2
\right)y
}{
1+2\langle x,y\rangle+\|x\|^2\|y\|^2
}.
\]
The additive inverse of $x$ under this operation is its Euclidean
negative, $-x$.

\subsection{Tangent spaces and geometric maps}
\label{app:geometric_maps}

The tangent space $T_z\mathbb{P}^{n}$ at any point
$z\in\mathbb{P}^{n}$ can be identified with $\mathbb{R}^{n}$ as
a vector space, although its inner product depends on $z$ through
the Riemannian metric.

The exponential map sends a tangent vector
$v\in T_z\mathbb{P}^{n}$ to a point on the manifold by following
the geodesic starting at $z$ in the direction $v$. As derived by \cite{ganea2018hyperbolic}, for $v\neq 0$,
it is given by
\[
\exp_z(v)
=
z\oplus
\left[
\tanh
\left(
\frac{\lambda_z\|v\|}{2}
\right)
\frac{v}{\|v\|}
\right],
\]
and
\[
\exp_z(0)=z.
\]

Conversely, the logarithmic map sends a point
$y\in\mathbb{P}^{n}$ to the tangent vector at $z$ that points
along the geodesic from $z$ to $y$. Let
\[
d=(-z)\oplus y.
\]
For $y\neq z$,
\[
\log_z(y)
=
\frac{2}{\lambda_z}
\operatorname{artanh}(\|d\|)
\frac{d}{\|d\|},
\]
and
\[
\log_z(z)=0.
\]

The Riemannian norm of this tangent vector equals the hyperbolic
distance:
\[
\left\|
\log_z(y)
\right\|_z
=
d_{\mathbb{P}}(z,y).
\]
Its ordinary Euclidean norm generally differs from the hyperbolic
distance because the tangent space metric is scaled by
$\lambda_z$.

At the origin, $\lambda_0=2$, and the maps reduce to
\[
\exp_0(v)
=
\begin{cases}
\tanh(\|v\|)
\dfrac{v}{\|v\|},
& v\neq 0,\\[0.6em]
0,
& v=0,
\end{cases}
\]
and
\[
\log_0(z)
=
\begin{cases}
\operatorname{artanh}(\|z\|)
\dfrac{z}{\|z\|},
& z\neq 0,\\[0.6em]
0,
& z=0.
\end{cases}
\]
These origin based expressions are used in the main text for
the tangent space analysis of continuous descriptor ratings.

\subsection{Parallel transport}
\label{app:parallel_transport}

Optimization on a Riemannian manifold requires comparing tangent vectors attached to different points. Since the tangent spaces $T_z\mathbb{P}^n$ and $T_y\mathbb{P}^n$ are distinct for $z \neq y$, tangent vectors cannot be directly added. The appropriate operation is \emph{parallel transport}, which moves a tangent vector along a geodesic while preserving the Riemannian geometry.

For $z,y\in\mathbb{P}^{n}$ and $v\in T_z\mathbb{P}^{n}$,
parallel transport along the geodesic from $z$ to $y$ is
\citep{ganea2018hyperbolic,becigneul2019riemannian}
\[
P_{z\rightarrow y}(v)
=
\frac{\lambda_z}{\lambda_y}
\operatorname{gyr}[y,-z]v.
\]
Here, the gyration operator associated with the gyrovector formalism of hyperbolic geometry is
defined by \citep{Ungar2009Gyrovector}:
\[
\operatorname{gyr}[a,b]v
=
-\left(a\oplus b\right)
\oplus
\left[
a\oplus\left(b\oplus v\right)
\right].
\]
Parallel transport is used during optimization to transfer the
first moment estimate between successive embedding positions.

\subsection{Embeddings initialization}
\label{app:initialization}

In order to initialize our embeddings, we need to sample points on the Poincaré disk. The hyperbolic space admits several generalizations of the Gaussian distribution, which are routinely deployed in statistical modeling and machine learning. For our purposes, we consider the \emph{pseudo-hyperbolic Gaussian} \citep{Nagano2019wrapped}. The latter is obtained by first sampling points $v$ on the tangent space $T_0\mathbb{P}^2$ according to a standard Gaussian distribution, and then projecting these points onto the hyperbolic disk: for each observation $i$, we sample $a_i
\sim
\mathcal{N}
\left(
0,
\sigma_{\mathrm{init}}^2 I_2
\right)$
in
$T_0\mathbb{P}^2,
$
and set $ z_i^{(0)} = \exp_0(a_i)$.
In other words, the pseudo-hyperbolic Gaussian is the push-forward via the exponential map of the standard Gaussian distribution over the tangent space at the origin. This procedure produces valid initial points inside the open unit disk. Independent samples are used for each random initialization. We set $\sigma_{\mathrm{init}}=0.1$ in all experiments.

\subsection{Riemannian Adam optimization}
\label{app:riemannian_adam}

To optimize the embedding coordinates, we employ a Riemannian Adam optimizer on the Poincaré disk, following the framework of \citet{becigneul2019riemannian} and instantiating it with the closed-form Poincaré expressions for the Riemannian gradient, exponential map, and parallel transport. Our implementation further uses the standard Adam bias correction \citep{kingma2015adam} and an explicit projection back into the open unit disk for numerical stability as performed by \cite{geoopt2020kochurov}.

Unlike the usual coordinate-wise form of Euclidean Adam, in our Riemannian implementation following \cite{becigneul2019riemannian}, each embedding point $z_i \in \mathbb{P}^2$ is associated with a single scalar second-moment estimate $v_i^{(t)}$, rather than separate second-moment estimates for its two coordinates. Let $z_i^{(t)} \in \mathbb{P}^2$ denote the embedding of observation $i$ at iteration $t$, and let $\nabla \mathcal{L}(z_i^{(t)}) \in \mathbb{R}^2$ be the corresponding Euclidean gradient of the loss $\mathcal{L}$ defined in the Section \ref{sec: hMDS}. The associated Riemannian gradient is
\[
g_i^{(t)}
=
\nabla_{\mathbb{P}} \mathcal{L}(z_i^{(t)})
=
\frac{(1-\|z_i^{(t)}\|^2)^2}{4}\,
\nabla \mathcal{L}(z_i^{(t)}).
\]
Let $\beta_1,\beta_2 \in [0,1)$ be hyperparameters controlling the exponential moving averages of the first-moment and second-moment estimates, respectively. We then maintain a first-moment estimate $m_i^{(t)} \in T_{z_i^{(t)}}\mathbb{P}^2$ and a scalar second-moment estimate $v_i^{(t)} \in \mathbb{R}_{+}$, updated as
\[
m_i^{(t)}
=
\beta_1 \,P_{z_i^{(t-1)}\rightarrow z_i^{(t)}}(m_i^{(t-1)})
+
(1-\beta_1)\, g_i^{(t)},
\]
\[
v_i^{(t)}
=
\beta_2 v_i^{(t-1)}
+
(1-\beta_2)\,
\|g_i^{(t)}\|_{z_i^{(t)}}^2,
\]
where $P_{z_i^{(t-1)}\rightarrow z_i^{(t)}}(m_i^{(t-1)})$ is the previous first moment transported to the current tangent space via \emph{parallel transport} from $T_{z_i^{(t-1)}}\mathbb{P}^2$ to $T_{z_i^{(t)}}\mathbb{P}^2$ in order to preserve the momentum across iterations, and
\[
\|g_i^{(t)}\|_{z_i^{(t)}}^2
=
g_{z_i^{(t)}}(g_i^{(t)},g_i^{(t)}).
\]
Following the standard Adam bias correction, we define
\[
\widehat{m}_i^{(t)}
=
\frac{m_i^{(t)}}{1-\beta_1^t},
\qquad
\widehat{v}_i^{(t)}
=
\frac{v_i^{(t)}}{1-\beta_2^t}.
\]
The tangent update direction is then
\[
h_i^{(t)}
=
-\eta \,
\frac{\widehat{m}_i^{(t)}}{\sqrt{\widehat{v}_i^{(t)}}+\varepsilon},
\]
where $\eta>0$ is the learning rate and $\varepsilon>0$ is a numerical stability constant. The embedding is updated intrinsically on the manifold via the exponential map:
\[
z_i^{(t+1)}
=
\exp_{z_i^{(t)}}\!\left(h_i^{(t)}\right).
\]

Finally, for numerical stability, we project points back into the open unit disk whenever needed. Specifically, after each update we apply
\[
\Pi_{\mathbb{P}^2}(u)
=
\begin{cases}
u, & \text{if } \|u\| < 1-\varepsilon_{\mathrm{proj}},\\[0.4em]
(1-\varepsilon_{\mathrm{proj}})\dfrac{u}{\|u\|}, & \text{otherwise}.,
\end{cases}
\]

\paragraph{Implementation details.} For all our experiments, we set $\varepsilon_{\mathrm{proj}} = 10^{-5}$, $\beta_1=0.9$, $\beta_2=0.999$, $\varepsilon=10^{-8}$, and $\eta=0.1$. For the Sagar dataset, optimization was performed in full
batch. Because of the larger size of the GSLF dataset, its embedding was optimized using mini-batches of 195 molecules. All configurations were trained for 1000 epochs. In each case, the optimization loss had reached a stable plateau by the end of training, indicating convergence.

\newpage
\section{Additional results}

\begin{table}[h]
\centering
\small
\begin{tabular}{lcccc}
\toprule
Descriptor & Radial Pearson & Radial Spearman & $p_{\mathrm{WS}}$ & $p_{\mathrm{OB}}$ \\
\midrule
Intensity & $0.42 \pm 0.05$ & $0.38 \pm 0.05$ & $<0.001$ & $<0.001$ \\
Pleasantness & $0.10 \pm 0.10$ & $0.07 \pm 0.10$ & $0.835$ & $0.873$ \\
Fishy & $0.02 \pm 0.07$ & $-0.14 \pm 0.06$ & $0.657$ & $0.665$ \\
Burnt & $-0.04 \pm 0.07$ & $-0.19 \pm 0.07$ & $0.356$ & $0.317$ \\
Sour & $0.05 \pm 0.06$ & $-0.08 \pm 0.05$ & $0.220$ & $0.207$ \\
Decayed & $0.08 \pm 0.07$ & $-0.11 \pm 0.07$ & $0.051$ & $0.044$ \\
Musky & $0.04 \pm 0.09$ & $-0.06 \pm 0.09$ & $0.966$ & $0.968$ \\
Fruity & $0.09 \pm 0.12$ & $-0.16 \pm 0.11$ & $0.049$ & $0.052$ \\
Sweaty & $0.09 \pm 0.09$ & $-0.08 \pm 0.09$ & $0.727$ & $0.755$ \\
Cool & $-0.11 \pm 0.09$ & $-0.22 \pm 0.09$ & $0.032$ & $0.029$ \\
Floral & $-0.12 \pm 0.12$ & $-0.24 \pm 0.11$ & $0.005$ & $0.003$ \\
Sweet & $0.10 \pm 0.13$ & $-0.02 \pm 0.13$ & $0.091$ & $0.167$ \\
Warm & $0.01 \pm 0.10$ & $-0.06 \pm 0.11$ & $0.952$ & $0.929$ \\
Bakery & $0.02 \pm 0.15$ & $-0.09 \pm 0.15$ & $0.813$ & $0.769$ \\
Spicy & $-0.17 \pm 0.03$ & $-0.30 \pm 0.03$ & $0.037$ & $0.058$ \\
\bottomrule
\end{tabular}
\caption{Full radial descriptor results for the Sagar union embedding. Radial Pearson and Spearman correlations quantify the association between each descriptor and the hyperbolic radius. Values are reported as mean and standard deviation over 10 random seeds. The restricted permutation p-values correspond to the within-subject permutation, $p_{\mathrm{WS}}$, and the odorant-block permutation, $p_{\mathrm{OB}}$.}
\label{tab:appendix_sagar_radial_all}
\end{table}

\begin{table}[h]
\centering
\small
\begin{tabular}{lccc}
\toprule
Descriptor & Directional $R^2$ & $p_{\mathrm{WS}}$ & $p_{\mathrm{OB}}$ \\
\midrule
Intensity & $0.04 \pm 0.05$ & $0.003$ & $0.005$ \\
Pleasantness & $0.40 \pm 0.09$ & $<0.001$ & $<0.001$ \\
Fishy & $0.23 \pm 0.04$ & $<0.001$ & $<0.001$ \\
Burnt & $0.19 \pm 0.05$ & $<0.001$ & $<0.001$ \\
Sour & $0.03 \pm 0.04$ & $<0.001$ & $<0.001$ \\
Decayed & $0.34 \pm 0.06$ & $<0.001$ & $<0.001$ \\
Musky & $0.51 \pm 0.06$ & $<0.001$ & $<0.001$ \\
Fruity & $0.46 \pm 0.09$ & $<0.001$ & $<0.001$ \\
Sweaty & $0.13 \pm 0.06$ & $<0.001$ & $<0.001$ \\
Cool & $0.12 \pm 0.10$ & $<0.001$ & $<0.001$ \\
Floral & $0.25 \pm 0.11$ & $<0.001$ & $<0.001$ \\
Sweet & $0.51 \pm 0.12$ & $<0.001$ & $<0.001$ \\
Warm & $0.26 \pm 0.12$ & $<0.001$ & $<0.001$ \\
Bakery & $0.25 \pm 0.13$ & $<0.001$ & $<0.001$ \\
Spicy & $0.05 \pm 0.02$ & $<0.001$ & $<0.001$ \\
\bottomrule
\end{tabular}
\caption{Full angular descriptor results for the Sagar union embedding. Directional $R^2$ quantifies how well each descriptor is explained by a global linear direction in the tangent-space representation of the Poincaré disk. Values are reported as mean and standard deviation over 10 random seeds. The restricted permutation p-values correspond to the within-subject permutation, $p_{\mathrm{WS}}$, and the odorant-block permutation, $p_{\mathrm{OB}}$.}
\label{tab:appendix_sagar_angular_all}
\end{table}

\begin{figure*}[p]
    \centering

    \begin{adjustbox}{
        max width=\textwidth,
        max totalheight=0.92\textheight,
        center
    }
    \begin{minipage}{\textwidth}
        \centering

        \begin{subfigure}[t]{0.31\linewidth}
            \centering
            \includegraphics[width=\linewidth]{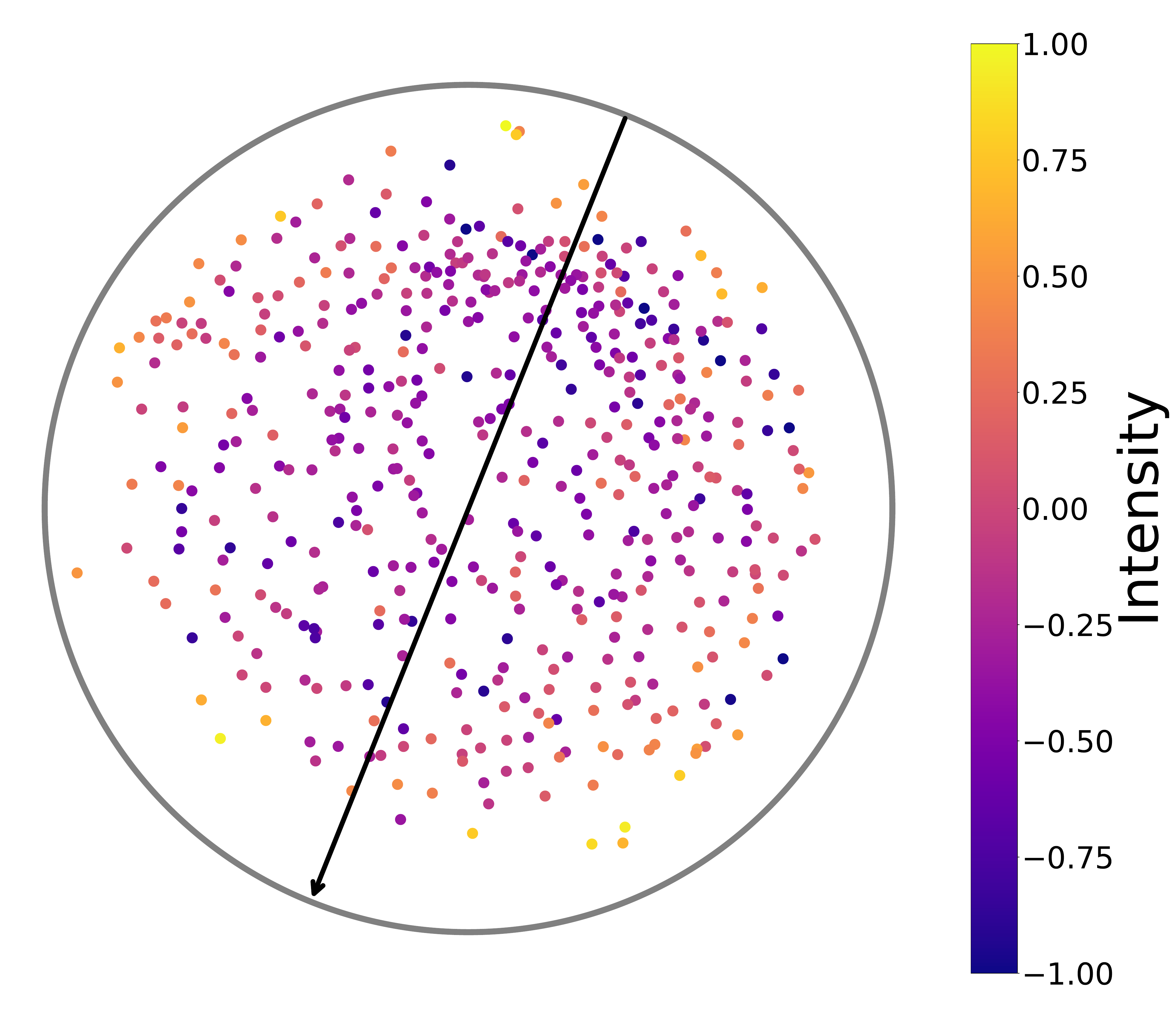}
            \caption{Intensity: $\theta=248.14^\circ$, $R^2=0.0150$.}
        \end{subfigure}
        \hfill
        \begin{subfigure}[t]{0.31\linewidth}
            \centering
            \includegraphics[width=\linewidth]{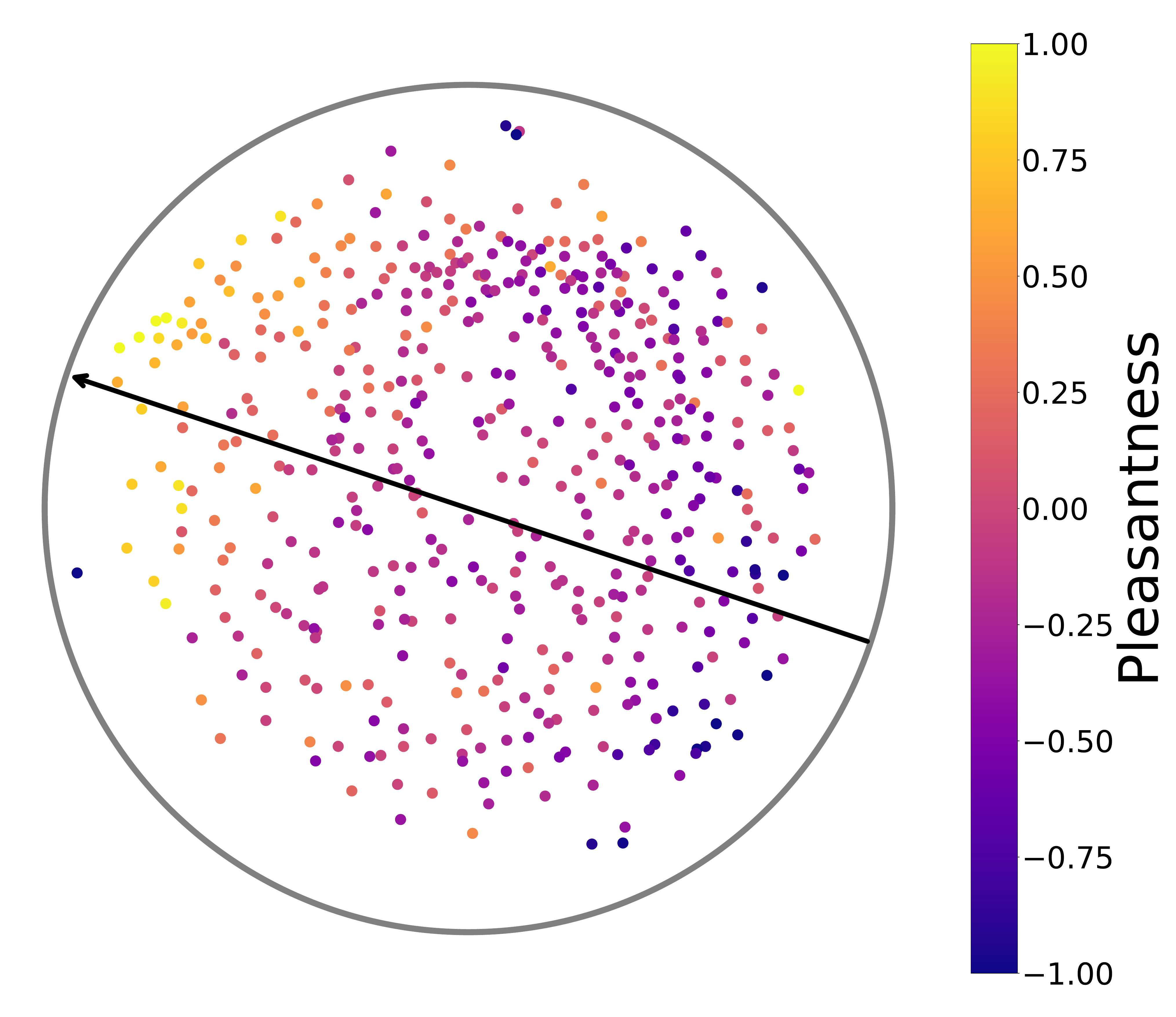}
            \caption{Pleasantness: $\theta=161.57^\circ$, $R^2=0.3736$.}
        \end{subfigure}
        \hfill
        \begin{subfigure}[t]{0.31\linewidth}
            \centering
            \includegraphics[width=\linewidth]{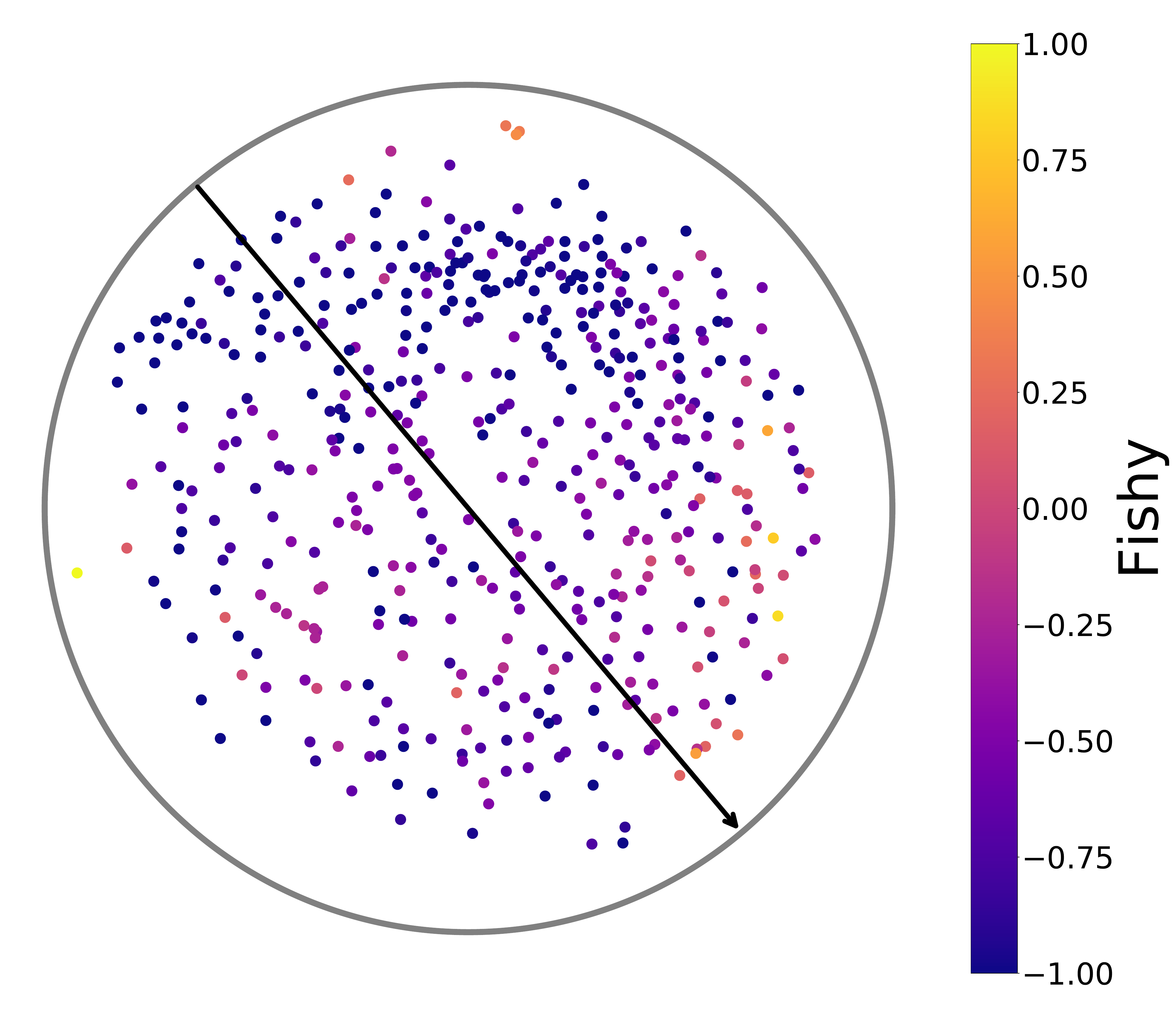}
            \caption{Fishy: $\theta=310.09^\circ$, $R^2=0.1442$.}
        \end{subfigure}

        \vspace{0.2em}

        \begin{subfigure}[t]{0.31\linewidth}
            \centering
            \includegraphics[width=\linewidth]{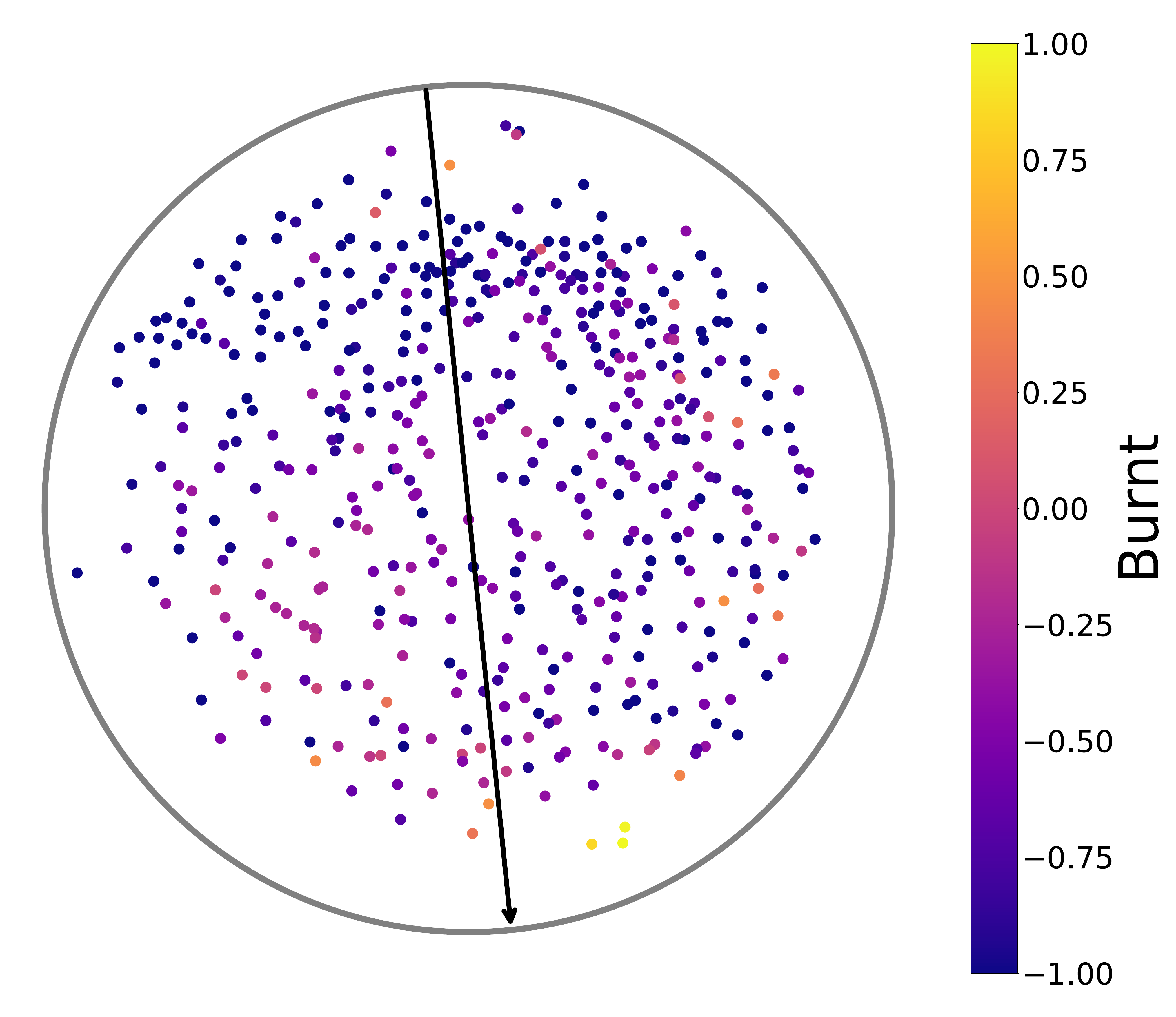}
            \caption{Burnt: $\theta=275.80^\circ$, $R^2=0.1786$.}
        \end{subfigure}
        \hfill
        \begin{subfigure}[t]{0.31\linewidth}
            \centering
            \includegraphics[width=\linewidth]{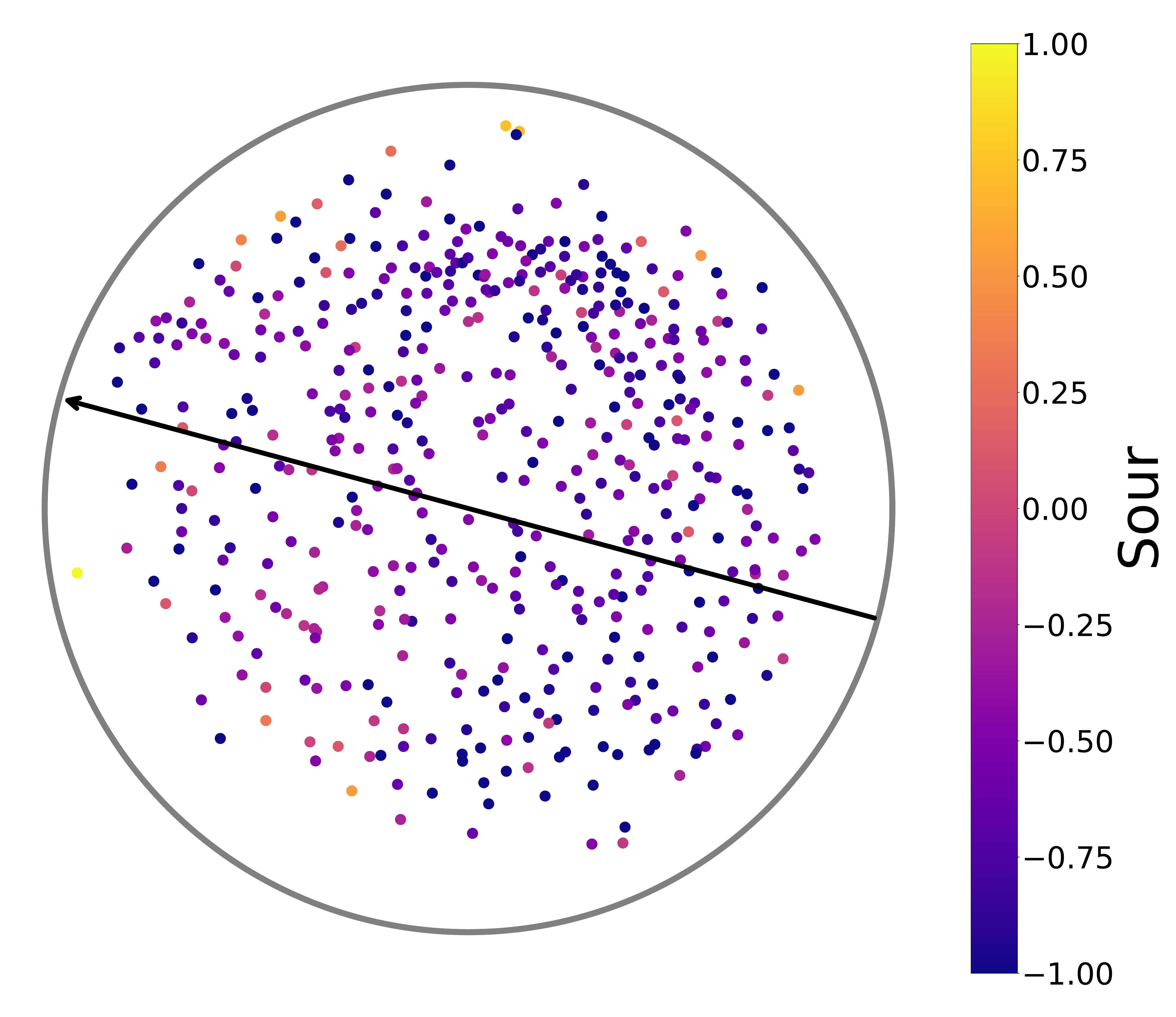}
            \caption{Sour: $\theta=164.89^\circ$, $R^2=0.0247$.}
        \end{subfigure}
        \hfill
        \begin{subfigure}[t]{0.31\linewidth}
            \centering
            \includegraphics[width=\linewidth]{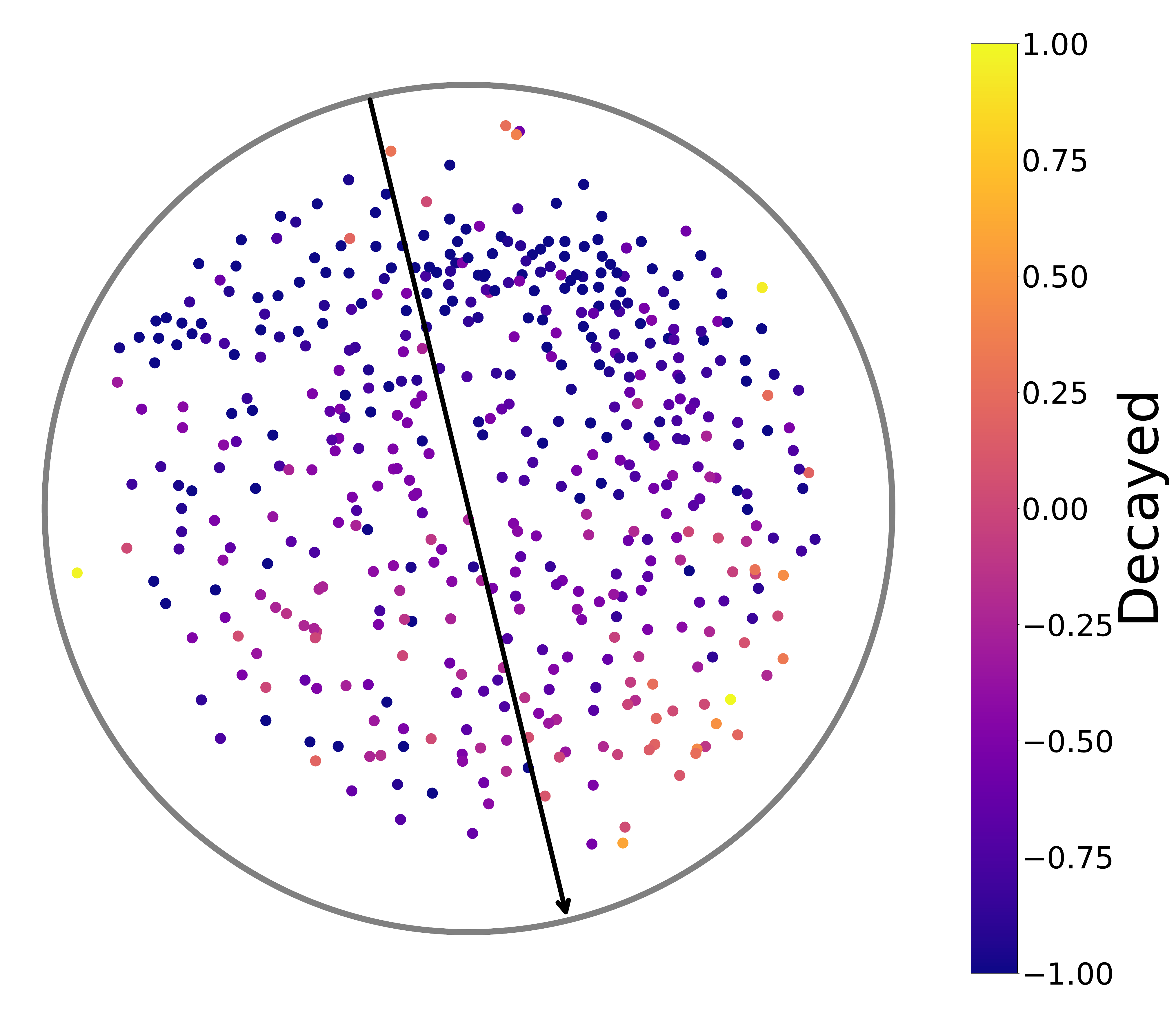}
            \caption{Decayed: $\theta=283.55^\circ$, $R^2=0.2608$.}
        \end{subfigure}

        \vspace{0.2em}

        \begin{subfigure}[t]{0.31\linewidth}
            \centering
            \includegraphics[width=\linewidth]{images/images_ChemicalSenses/1000_embeddings_6.pdf}
            \caption{Musky: $\theta=279.97^\circ$, $R^2=0.4291$.}
        \end{subfigure}
        \hfill
        \begin{subfigure}[t]{0.31\linewidth}
            \centering
            \includegraphics[width=\linewidth]{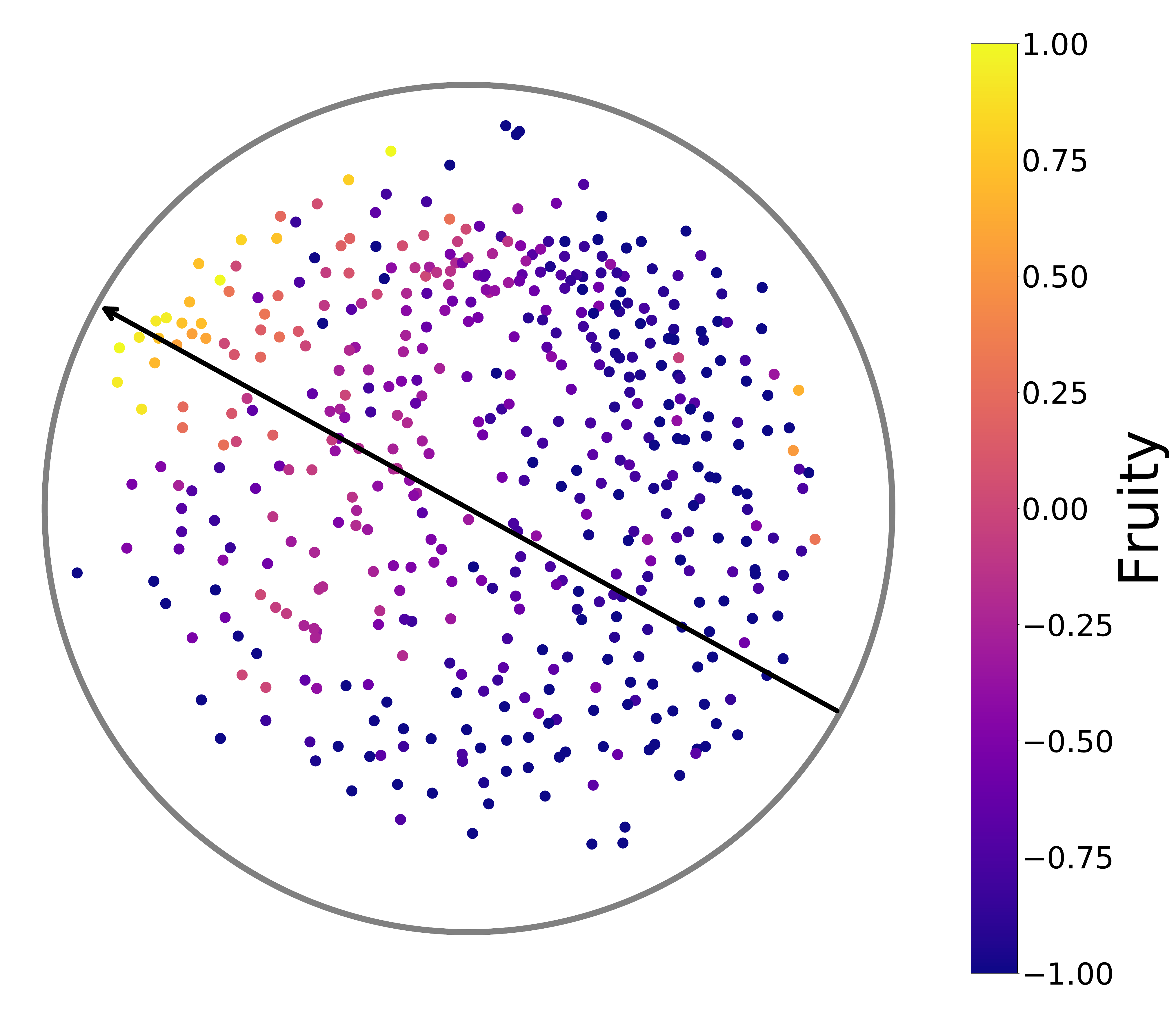}
            \caption{Fruity: $\theta=151.22^\circ$, $R^2=0.4455$.}
        \end{subfigure}
        \hfill
        \begin{subfigure}[t]{0.31\linewidth}
            \centering
            \includegraphics[width=\linewidth]{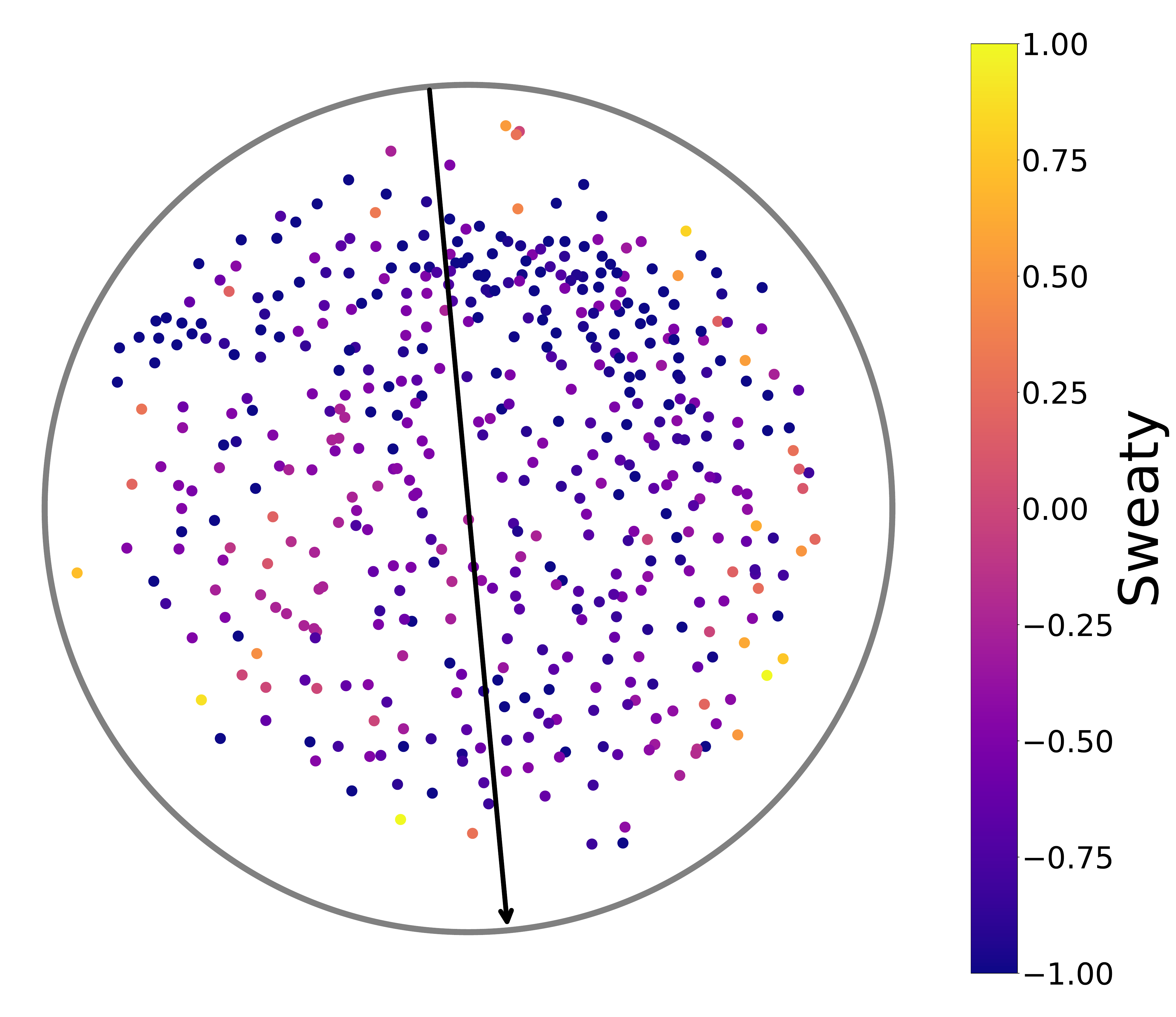}
            \caption{Sweaty: $\theta=275.33^\circ$, $R^2=0.0522$.}
        \end{subfigure}

        \vspace{0.2em}

        \begin{subfigure}[t]{0.31\linewidth}
            \centering
            \includegraphics[width=\linewidth]{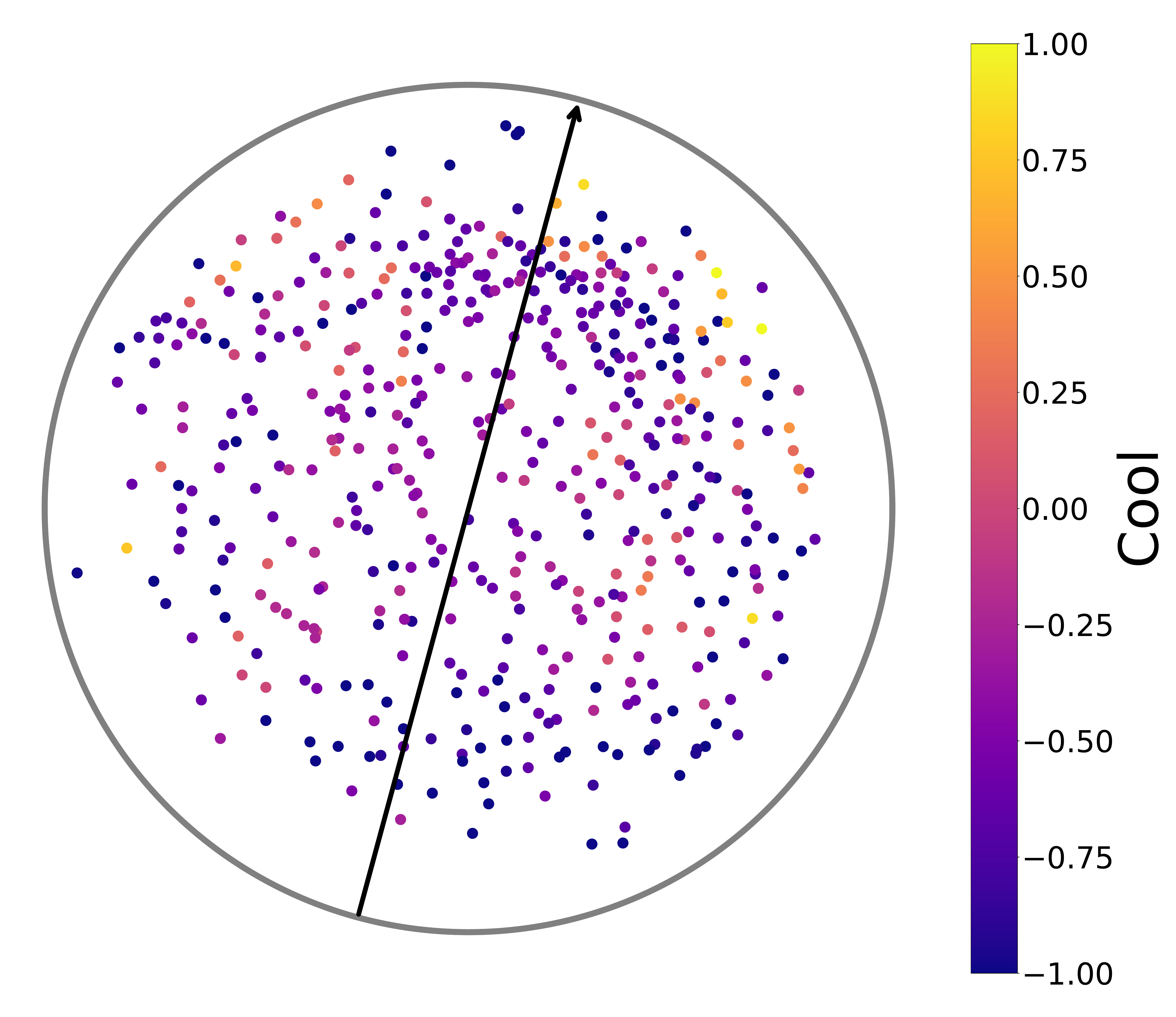}
            \caption{Cool: $\theta=74.85^\circ$, $R^2=0.0427$.}
        \end{subfigure}
        \hfill
        \begin{subfigure}[t]{0.31\linewidth}
            \centering
            \includegraphics[width=\linewidth]{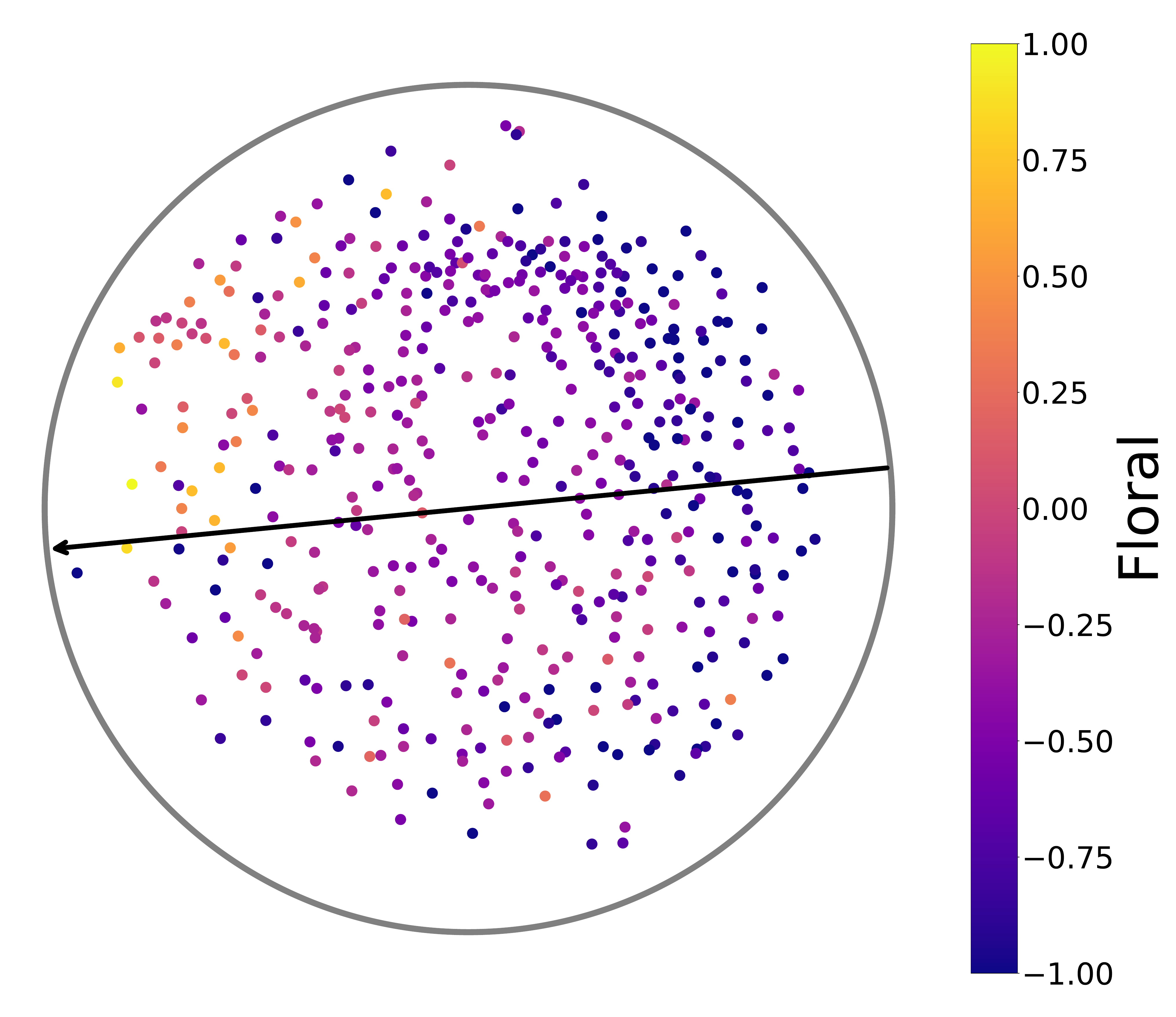}
            \caption{Floral: $\theta=185.54^\circ$, $R^2=0.3299$.}
        \end{subfigure}
        \hfill
        \begin{subfigure}[t]{0.31\linewidth}
            \centering
            \includegraphics[width=\linewidth]{images/images_ChemicalSenses/1000_embeddings_11.pdf}
            \caption{Sweet: $\theta=158.40^\circ$, $R^2=0.5828$.}
        \end{subfigure}

        \vspace{0.2em}

        \begin{subfigure}[t]{0.31\linewidth}
            \centering
            \includegraphics[width=\linewidth]{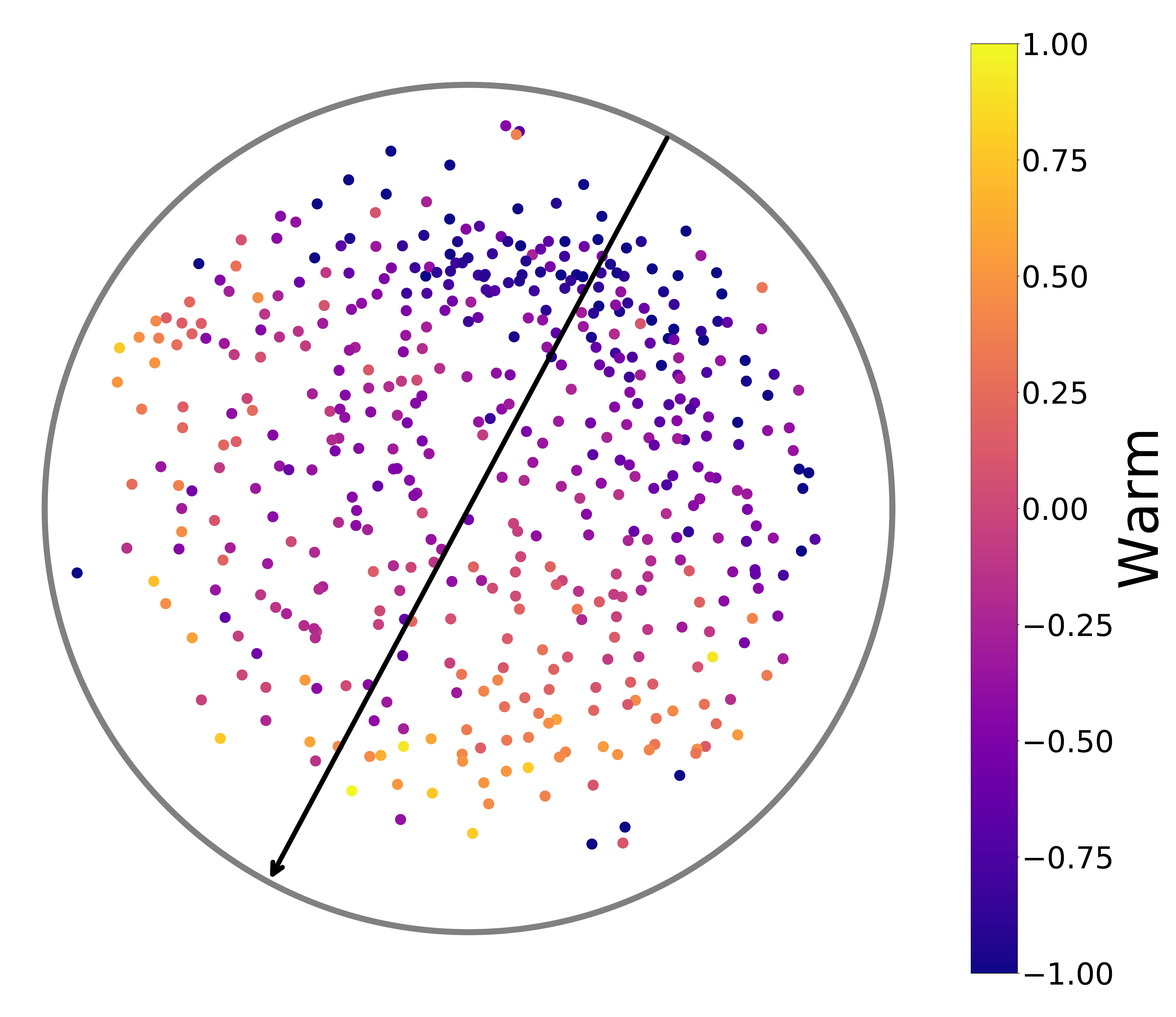}
            \caption{Warm: $\theta=241.83^\circ$, $R^2=0.4636$.}
        \end{subfigure}
        \hfill
        \begin{subfigure}[t]{0.31\linewidth}
            \centering
            \includegraphics[width=\linewidth]{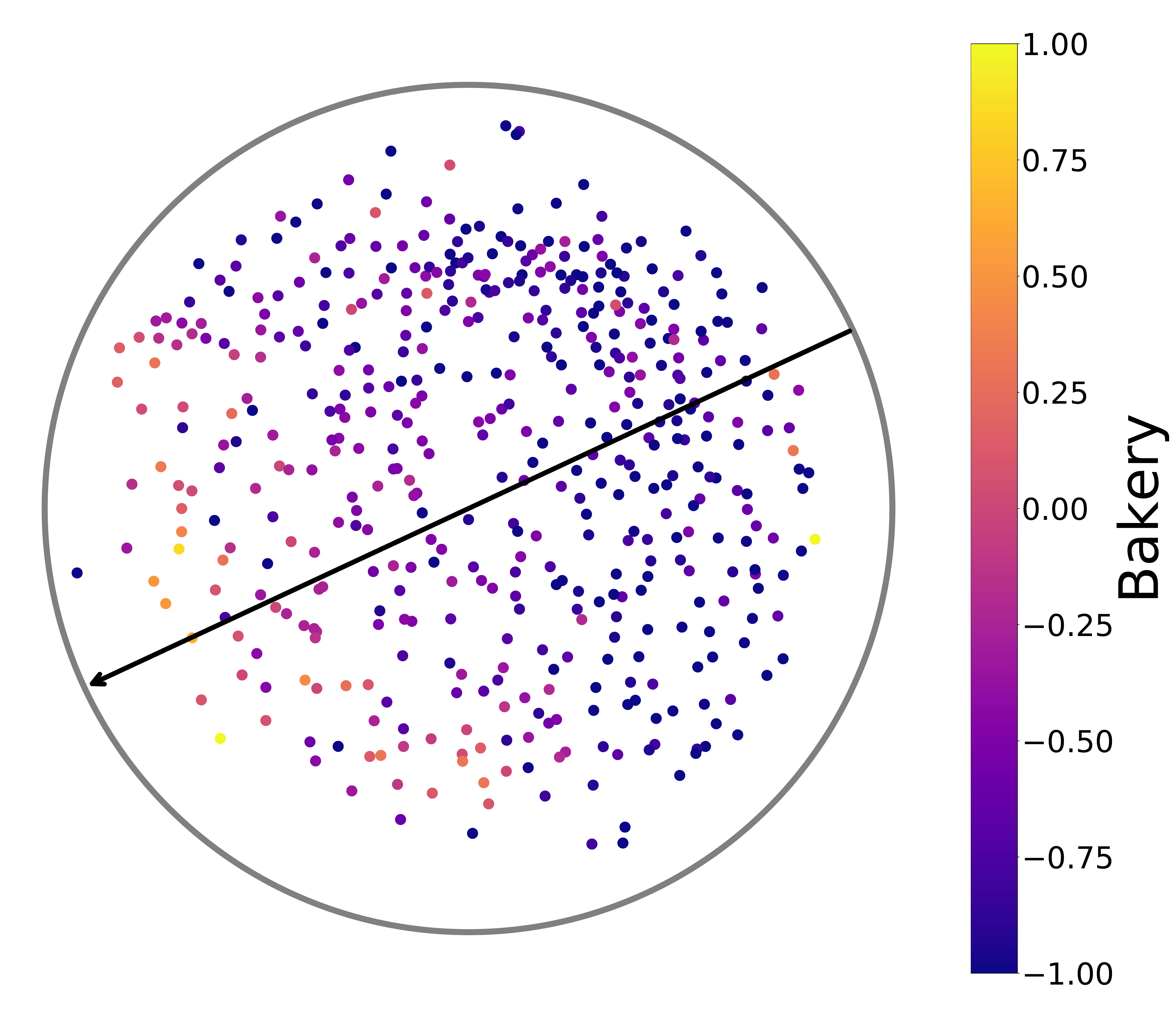}
            \caption{Bakery: $\theta=204.98^\circ$, $R^2=0.2961$.}
        \end{subfigure}
        \hfill
        \begin{subfigure}[t]{0.31\linewidth}
            \centering
            \includegraphics[width=\linewidth]{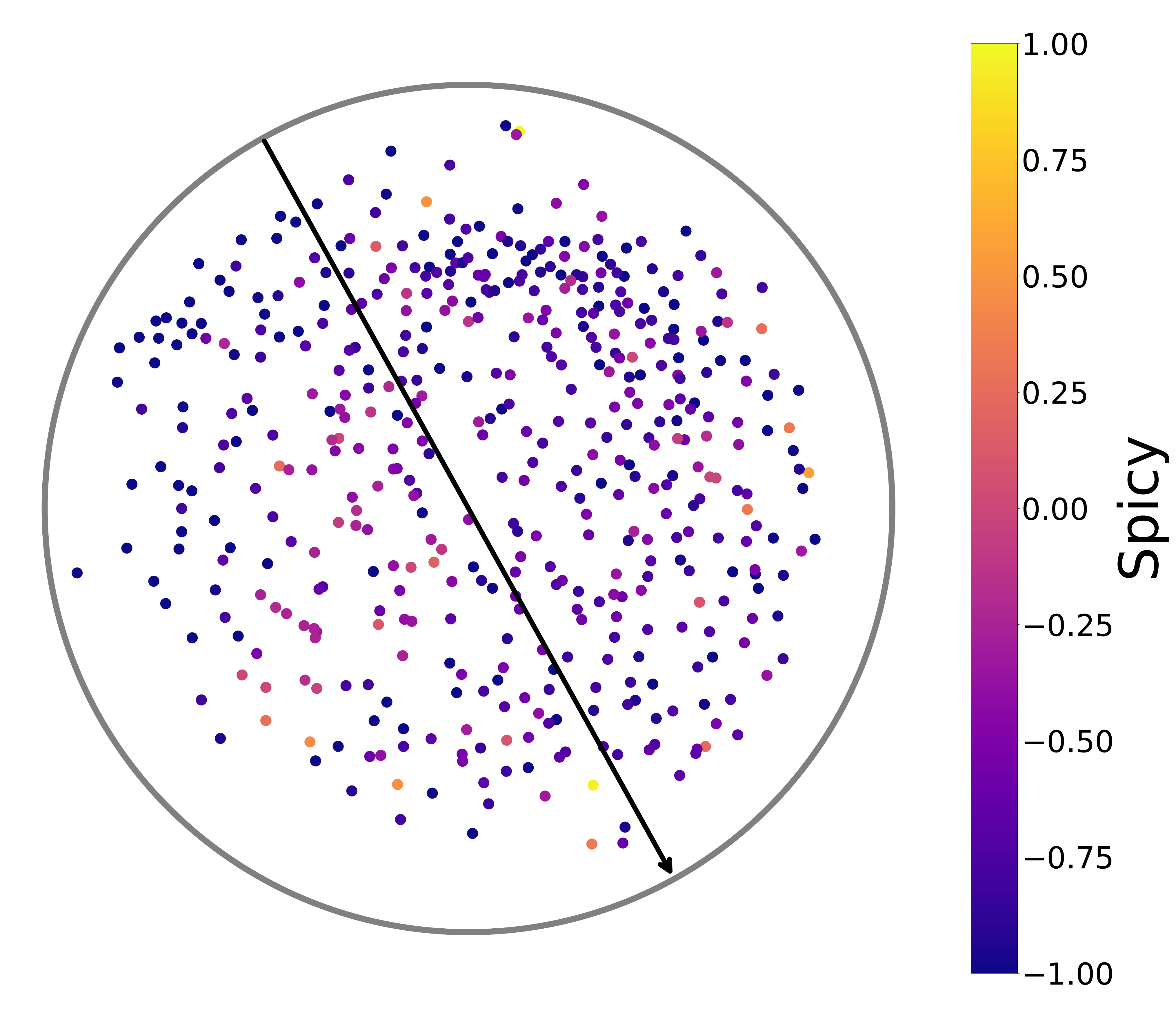}
            \caption{Spicy: $\theta=299.05^\circ$, $R^2=0.0285$.}
        \end{subfigure}

    \end{minipage}
    \end{adjustbox}

    \caption{
    Directional organization of the 15 continuous descriptors
    in a representative Sagar union embedding. Each panel shows
    the embedding colored by the corresponding descriptor rating.
    Arrows indicate the fitted tangent space directions of
    increasing descriptor values. Here, $\theta$ denotes the
    directional angle, and $R^2$ quantifies the strength of the
    fitted directional trend. The reported values are specific
    to the random initialization ($m=5$).\\
    \textbf{Alt text:} Fifteen Poincaré disk panels show directional organization for each Sagar descriptor.
    }
    \label{fig:all_descriptor_directions}
\end{figure*}

\end{document}